\documentclass[useAMS,usenatbib]{mn2e}
\usepackage{graphicx}
\usepackage{latexsym}
\usepackage{amssymb}

\title[{Satellite profiles in groups and clusters}]{The radial distribution of galaxies in groups and clusters}

\author[J. M. Budzynski et al.]{J.~M.~Budzynski,$^1$\thanks{E-mail: jbudzyn@ast.cam.ac.uk} S.~Koposov,$^{1,2}$ I.~G.~McCarthy$^{3,4}$, S.~L.~McGee$^5$
\newauthor{\& V.~Belokurov$^1$} \\
$^{1}$Institute of Astronomy, University of Cambridge, Madingley Road, Cambridge, CB3 OHA\\
$^2$Sternberg Astronomical Institute, Moscow State University, Universitetskiy pr. 13, 119992 Moscow, Russia \\
$^{3}$Kavli Institute for Cosmology, University of Cambridge, Madingley Road, Cambridge, CB3 OHA\\
$^4$Astrophysics and Space Research Group, School of Physics and Astronomy,
University of Birmingham, Edgbaston, Birmingham B15 2TT\\
$^5$Institute for Computational Cosmology, University of Durham, South Road, Durham, DH1 3LE
\\}

\begin{document}

\pagerange{\pageref{firstpage}--\pageref{lastpage}} \pubyear{2002}

\newcommand{\dd}{\textrm{d}}
\newcommand{\rin}{R_\textrm{\tiny{in}}}
\newcommand{\OO}{\mathcal{O}}
\maketitle

\label{firstpage}

\def\mnras{MNRAS}
\def\apj{ApJ}
\def\aap{A\&A}
\def\aaps{A\&A}
\def\apjl{ApJL}
\def\aj{AJ}
\def\apjs{ApJS}
\def\araa{ARA\&A}
\def\nat{Nature}
\def\physrep{Physics Reports}
\def\na{New A}

\newcommand{\subfind}    {\textsc{Subfind}}
\newcommand{\basenum}{55\,121}
\newcommand{\finsample}{20\,157}
\newcommand{\numclusttot}{129}

\begin{abstract}
We present a new catalogue of \basenum \ groups and clusters centred on Luminous Red Galaxies from SDSS DR7 in the redshift range $0.15\leq z \leq 0.4$. We provide halo mass ($M_{500}$) estimates for each of these groups derived from a calibration between the optical richness of bright galaxies ($M_{r}\leq-20.5$) within 1 Mpc, and X-ray-derived mass for a small subset of \numclusttot \ groups and clusters with X-ray measurements. For \finsample \ high-mass groups and clusters with $M_{500}>10^{13.7}$ $M_{\odot}$, we find that the catalogue has a purity of $>97$ per cent and a completeness of $\sim 90$ per cent. We derive the mean (stacked) surface number density profiles of galaxies as a function of total halo mass in different mass bins.  We find that derived profiles can be well-described by a projected NFW profile with a concentration parameter ($\left<c\right> \equiv \left<r_{200}/r_s\right>\approx2.6$) that is approximately a factor of two lower than that of the dark matter (as predicted by N-body cosmological simulations) and nearly independent of halo mass.  Interestingly, in spite of the difference in shape between the galaxy and dark matter radial distributions, both exhibit a high degree of self-similarity.  We also stack the satellite profiles based on other observables, namely redshift, BCG luminosity, and satellite luminosity and colour.  We see no evidence for strong variation in profile shape with redshift over the range we probe or with BCG luminosity (or BCG luminosity fraction), but we do find a strong dependence on satellite luminosity and colours, in agreement with previous studies. A self-consistent comparison to several recent semi-analytic models of galaxy formation indicates that: (1) beyond $\approx 0.3 r_{500}$ current models are able to reproduce both the shape and normalisation of the satellite profiles; and (2) within $\approx 0.3 r_{500}$ the predicted profiles are sensitive to the details of the satellite-BCG merger timescale calculation.  The former is a direct result of the models being tuned to match the global galaxy luminosity function combined with the assumption that the satellite galaxies do not suffer significant tidal stripping, even though their surrounding dark matter haloes can be removed through this process.  Combining our results with measurements of the intracluster light should provide a way to inform theoretical models on the efficacy of the tidal stripping and merging processes.

\end{abstract}


\section{Introduction} 

It has been recognized for some time now that the abundance and radial distribution of satellite galaxies within larger host systems
can potentially provide strong tests of our current cosmological
paradigm for structure formation. Perhaps the most well-known example of this is the so-called `missing satellites problem' of cold dark
matter (CDM) models, which refers to the apparent inconsistency between the CDM predictions and the Local 
Group satellite census. As pointed out by several groups \citep[e.g.][]{1999ApJ...524L..19M, 1999ApJ...522...82K}, the count 
of the Milky Way and M31 satellites is lagging behind the cumulative sub-halo mass function by several orders 
of magnitude. The solution to this problem may simply be that galaxy formation becomes
very inefficient at low halo masses (perhaps due to feedback processes
such as photoionisation and winds from supernovae; e.g., \citealt{2000ApJ...539..517B, 2002ApJ...572L..23S, 2004ApJ...609..482K, 2009ApJ...696.2179K}), or it may signal a
more fundamental problem with the underlying CDM theoretical
framework \citep{2000PhRvL..84.4525K, 2000PhRvL..84.3760S, 2001ApJ...556...93B, 2003ApJ...598...49Z}.

The abundance and distribution of satellite galaxies in more massive
galaxy groups and clusters (hereafter collectively referred to as clusters) provides astronomers with
another valuable check of our current structure formation paradigm.
Furthermore, as can be readily shown with simple `abundance matching'
\citep{2004MNRAS.353..189V,2006ApJ...647..201C, 2006ApJ...643...14S, 2007ApJ...668..826C, 2008MNRAS.388..945B, 2010ApJ...710..903M, 2010MNRAS.404.1111G} arguments\footnote{A procedure to link galaxies to their DM haloes by matching their observed stellar mass functions to simulated halo mass functions.}, a sizeable
fraction of the overall galaxy population resides in (i.e., are
satellites of) clusters.  The processes that influence the abundance and
distribution of satellites in clusters must therefore be properly taken
into consideration in the development of a general theory for galaxy
formation.  These processes can be constrained by directly
confronting observations of the radial distribution of satellites with
theoretical predictions.  Finally, the formation and evolution of the
most massive galaxies in the universe, i.e., the central brightest
cluster galaxies (BCGs), as well as that of the diffuse intracluster
light\footnote{Which, by some estimates, may contain as much as 50\%
  of the integrated stellar mass of clusters \citep[e.g.][]{2007ApJ...666..147G}.} (ICL), are currently believed to be intimately linked
(via mergers and tidal disruption) to the satellite galaxy population.
A detailed understanding of BCGs and the ICL therefore necessarily
includes an understanding of the satellite population.

The advent of large optical surveys and near-IR surveys has allowed
for more detailed characterisation of radial profiles of satellites in
clusters than was previously possible.  \citet[][see also \citeauthor{2000AJ....119.2038V} \ \citeyear{2000AJ....119.2038V}]{1997ApJ...478..462C}
and \citet{2007ApJ...659.1106M} measured the
r-band and K-band (respectively) satellite number density profiles of
15 X-ray-selected galaxy clusters ($z \sim 0.3$) in the CNOC1 survey.
Both studies concluded the radial profiles were consistent with
relatively cuspy distributions (in the inner regions) but with a
concentration ($c \equiv r_{200}/r_s \approx 4$, where $r_{200}$ is
the radius that encloses a mean density of 200 times the critical
density of Universe and $r_s$ is the fitted scale radius) that is
somewhat lower than that of the underlying dark matter distribution
predicted from dissipationless cosmological simulations.  Using the
satellites themselves as tracers of the mass, \citeauthor{1997ApJ...478..462C} performed a Jeans analaysis to measure the total mass-to-light
ratio as a function of radius and indeed found that beyond
approximately $0.3 r_{200}$ the satellites/light traced the underlying
mass distribution well.  \citet{2004ApJ...610..745L} stacked a sample of 93
local ($z \approx 0.05$) X-ray-selected galaxy clusters observed with
2MASS to derive the mean projected K-band satellite number density
profile.  Consistent with the previously mentioned studies, \citet{2004ApJ...610..745L} found the stacked satellite number density profile could be well
fit by a NFW distribution, but with a somewhat lower concentration
still, of $c \approx 2.9$.  The difference in the derived
concentrations may reflect differences in the way $r_{200}$ was
estimated (\citeauthor{1997ApJ...478..462C} use velocity-dispersion based method,
whereas \citeauthor{2004ApJ...610..745L} used a X-ray temperature-mass scaling relation),
differences sample selection, and differences in the mean redshifts of
the samples.  Due to the relatively small sample sizes and the limited
mass range studied, \citet{1997ApJ...478..462C} and \citet{2004ApJ...610..745L}
were unable to explore any dependencies the radial profiles may have
on host system mass.

Somewhat more recently, \citet{2005ApJ...633..122H} used the original maxBCG
cluster sample derived from the SDSS to investigate the spatial
distribution of satellites.  The maxBCG algorithm finds overdensities
of galaxies which have colours that place them on the red sequence of
galaxies and which have a BCG that has colours that are compatible
with the red sequence.  The large survey area affored by the SDSS
meant \citeauthor{2005ApJ...633..122H} could study a much larger sample of clusters than was
previously possible.  They studied a sample of 6708 clusters in total and
derived the stacked number density profiles of satellites in several
bins of richness.  Consistent with the previous findings,
\citeauthor{2005ApJ...633..122H} found relatively cuspy distributions of the satellites.
Interestingly, the inferred concentrations were much lower (they found
$c \approx 1$) than that inferred previously for high mass clusters
and, furthermore, found a relatively steep trend in concentration with
cluster richness.  However, it is important to note that \citeauthor{2005ApJ...633..122H}
adopted a different definition for the concentration than
the previously mentioned studies.  In particular, lacking a robust
total halo mass estimate of the clusters in their sample, they adopted
a more observationally-motivated definition for $r_{200}$ (which they
label as $R_{200}^{N}$), which they define as the radius at which the
space density of galaxies is overdense by a factor of $200$.  The
advantage of this choice is that the characteristic radius can be
measured directly from the data.  The disadvantage is that it
complicates comparisons to theoretical models and simulations.

On the theory side, the radial distribution of dark matter subhaloes
in dissipationless cosmological simulations has been well quantified
\citep[e.g.,][]{2001NewA....6...79S, 2008MNRAS.387..536G, 2004ApJ...609...35K, 2007MNRAS.382.1901S, 2007ApJ...659.1082S, 2009MNRAS.399..983A, 2010arXiv1002.3660K}.  In general, these studies find that at intermediate/large
cluster-centric radii ($r \ga 0.3-0.4 r_{200}$) the dark matter subhaloes
trace that of the underlying main halo.  At smaller radii, however,
the distribution of subhaloes flattens significantly (unlike observed
satellite galaxy profiles), which is most likely the result of
efficient tidal disruption of the subhaloes as they pass close to the
centre.  Recent semi-analytic models of galaxy formation \citep[e.g.][]{2006MNRAS.365...11C, 2006MNRAS.370..645B, 2007MNRAS.375....2D}, which are
based on merger trees extracted from such dissipationless cosmological
simulations, by necessity have to adopt assumptions about the
dynamical evolution of stellar component of the satellites galaxies
which is not included in the dissipationless simulations.  In general,
the derived satellite galaxy density profiles are cuspier than the
subhalo distribution \citep[e.g.][]{2007MNRAS.382.1901S} and in better apparent
agreement with the observations, although to our knowledge no detailed
comparisons have been made between current semi-analytic models and
observations.  We address this point further below.

A number of recent numerical studies have attempted to predict the
radial distribution of satellites using self-consistent cosmological
hydrodynamic simulations \citep[e.g.][]{2005ApJ...618..557N,2006MNRAS.373..397S,2009MNRAS.399..497D}.  
In general, the predicted satellite
distributions are more cuspy than the distribution of subhaloes in
dissipationless simulations, owing to the increased resiliency (from a
tightly bound stellar component) of the satellites to tidal
disruption.  \citet{2005ApJ...618..557N} compared their simulations to the
observed profiles of \citet{1997ApJ...478..462C} and \citet{2004ApJ...610..745L}
and concluded that the simulated profiles approximately match the shape
of the observed profiles over the wide radial range of $0.1 \la
r_{200} \la 1.0$.  However, they did not compare the observed and
predicted normalisation of the profiles and due to the heavy
computational expense could only simulate a handful of systems and
therefore were unable to explore any depencence on host halo mass.

The main aim of the present study is to improve upon previous
observational measurements of the satellite number density profiles of
clusters.  To this end, we use the DR7 release of the SDSS.  Clusters
are indentified by looking for overdensities in fields centered on
Luminous Red Galaxies (LRGs, see Section 2).  Our final sample (which is publicly available\footnote{http://www.ast.cam.ac.uk/ioa/research/cassowary/lrg\textunderscore clusters \\ 
/lrg\textunderscore clusters\textunderscore dr7.fits}), after
all cuts, is \basenum, which is a factor of $\sim 8$ larger than \citet{2005ApJ...633..122H}. 
To normalise our radial profiles and to assign halo
masses for stacking, we use an optical richness-X-ray
temperature-total mass scaling relationship, to help facilitate
comparisons with models.  As we will show, we find that the derived
satellite profiles are relatively well-described by cuspy NFW distributions, as
found previously, but with a somewhat lower concentration, which is
roughly a factor of 2 lower than that of the dark matter distribution
(predicted from simulations).  Furthermore, unlike \citet{2005ApJ...633..122H},
we find no significant variation in the mean
concentration as a function of halo mass, which we mainly attribute to
the different definitions of $r_{200}$ that have been adopted (see
Section 5.2).  We also explore the dependence of the satellite radial distribution on several other properties, namely the redshift of the cluster, the luminosity of the BCG, and the luminosities and colours of the satellites themselves.

A secondary aim of the present paper is to make detailed comparisons
of the predictions of current semi-analytic models of galaxy formation
to our observations.  To that end, we use the public SQL database for
the Durham and Munich semi-analytic models to derive the predicted
satellite number density profiles.  We analyse the mock observations
in an identical way to the data (e.g., statistically background
subtract using random patches of the simulated sky and stack clusters in
an identical manner).  At intermediate/large radii, where the effects
of tidal disruption are evidently small, we find that the current models
reproduce our observed profiles remarkably well in both shape and
normalisation, the latter owing to the fact that the models have been
tuned to match the global galaxy luminosity function.  At smaller
radii, the results are sensitive to the details of how the merger
timescale for infalling satellites is calculated in the models.  Our
observations can therefore be used to inform the models on the
efficiency of satellite merrging and disruption.

The present paper is organised as follows.  In Section 2, we define the cluster sample and we describe how they are assigned an estimate of halo mass using a mass-richness X-ray calibration in Section 3. We also discuss the quality of the catalogue in terms of purity and completeness in Section 3. In Sections 4 \& 5, we present the construction of stacked number density profiles of the clusters in our sample, and investigate how the satellite concentration varies with halo mass and other observable quantities. Finally, in Section 6, we compare the observed satellite profiles to a series of semi-analytic models of galaxy formation.

Throughout this work we assume a cosmology of $\Omega _{\mathrm{M}}=0.27$, $\Omega _{\mathrm{\Lambda}}=0.73$, and $h=0.71$.

\section{Cluster sample}\label{sec:samples}

There are multiple strategies one can follow to create catalogues of
clusters from the data provided by the SDSS. A quick look at
Table 1 of \citet{2010ApJS..191..254H} reveals a growing family of
methods utilizing deep SDSS multi-band photometry to partition
galaxies in groups around one BCG
\citep{2007ApJ...660..239K, 2009ApJS..183..197W, 2010ApJS..191..254H}.
These algorithms proceed by first locating candidate BCGs and then
assembling a list of surrounding candidate satellites. This can be
done either in pseudo-3D space composed of two angular coordinates and
the photometric redshift (e.g., \citealt{2009ApJS..183..197W}) or by requiring
that the angular proximity is complemented by similarity in colour
(e.g., \citealt{2007ApJ...660..239K, 2010ApJS..191..254H}). The success of these
methods rests on the powerful assumption of existence of a BCG near
the bottom of the cluster's potential well.

We search for clusters in fields centered on LRG galaxies (i.e., each LRG is a potential BCG sitting at the center of a cluster).
Not only does this simplify and speed up the identification of clusters, it improves the
reliability of the search in the higher redshift range where we do not
have to rely heavily on increasingly more uncertain photometric
redshift. With this simple modification, we can now take advantage of
the large number of LRGs available in the SDSS DR7. To select the
LRGs, we apply the standard photometric cuts
\citep{2001AJ....122.2267E, 2011arXiv1103.2700T}, which are designed
to allow for a passively evolving stellar population. Specifically, we
shall use the colour, magnitude and surface brightness criteria set
down in equations 4-13 of \citet{2001AJ....122.2267E}. 

We restrict our clusters to lie in the redshift $0.15 \le z
\leq 0.4$. The lower limit is in place to minimise photometric
contamination of the LRG sample \citep{2010MNRAS.405.2534T}, and the
upper redshift limit allows for an approximately volume limited sample of LRGs and a complete sample of satellite
galaxies down to a modest magnitude limit $M_{r}=-20.5$ for all
galaxies with $z \leq 0.4$ (See Section \ref{sec:virmass} for
details). Although DR7 LRGs are not perfectly sampled due to fibre collision effects, we do not expect our results to be affected as their completeness is $\geq$95 per cent \citep{2001AJ....122.2267E}. 
We find $\sim 85\,000$ LRGs in the above redshift range in
SDSS DR7, which we further reduce by $\sim 3,000$ by eliminating
fields containing gaps or survey edges (i.e. with less than 85 per cent area completeness per field).

Our final cut is to ensure that there exists only one BCG per system.  Recall that our approach is assume that all LRGs are potential BCGs living at the centers of clusters.  In reality, some clusters (particularly massive clusters) will contain more than a single LRG and by default these systems would be identified as separate clusters, which is obviously undesirable.  To address this issue, we perform a LRG neighbour search within $r_{200}$ in angular separation and $\Delta z_{\mathrm{LRG}}\leq0.02$, from the current LRG and keep the brightest one as the BCG. Any other LRGs within the cluster aperture are kept as satellites. This removes a further $\sim 10\,000$ duplicate objects from the sample and leaves a 
sample of $\sim 72\,000$ cluster candidates. Although findings by \citet{2010AAS...21533002S} suggest that the central galaxy in the cluster may not always be the brightest, we investigate the effects of cluster mis-centering in Section \ref{sub:miscentre}, and find that these effects are small and do not affect our results. We find that a large fraction of these $70\,000$ ($\sim 20$ per cent) cluster candidates
in fact contain no excess galaxy overdensity (or richness above the background) in the surrounding environment (see Section \ref{sec:optobs} for details). These `field' systems are correspondingly removed from our sample, which yields a final base sample of
\basenum \ clusters. This number exceeds
the latest sample published by \citet{2009ApJS..183..197W} by a factor
of 2 and is similar in size to the GMBCG catalogue
\citep{2010ApJS..191..254H}. Naturally, owing to the similarity of the
search techniques, these cluster catalogs have a large number of objects in
common. We provide the interested reader with a further comparison in
the later sections.

In what follows, when calculating the stacked radial satellite
profiles of our clusters, we will assume the clusters' centers to lie
at the BCG position. There is, of course, some doubt as to how big the
offset between the true centre and the BCG could be. We investigate
this uncertainty in the Section \ref{sub:miscentre}. Note that not only our approach
is convenient, but it provides an accurate and well-defined stacking
centroid centre that can be used in the future for stacking of galaxy
agglomerations of any apparent richness.

Throughout the analysis presented in this paper, we use magnitudes
corrected for effects of dust extinction (Schlegel et al).  BCG
absolute magnitudes are calculated using K-corrections provided by the
calculation package provided by \citet{2010MNRAS.405.1409C}. The \citeauthor{2010MNRAS.405.1409C} polynomial fits provide no correction for passive evolution, and we do not make use of any additional evolution corrections in this work. The reason for this is three-fold. Firstly, in this work we calculate the absolute magnitude of both satellite galaxies and LRGs, and thus we want to avoid any additional uncertainty in the evolution corrections due to uncertain stellar populations. Secondly, an important part of this work is a comparison of the observations to various semi-analytic models, whose synthetic galaxy magnitudes also contain no evolution corrections. Finally, by applying evolution corrections to our central LRGs and redoing the analysis as a test, we find that our results are unchanged.

A summary of the mean brightness and colour properties of the LRG and satellite populations in our sample is shown in Table \ref{tab:brightscols}. As expected the central LRGs are significantly brighter and redder than the surrounding satellite galaxy populations.

\begin{table}
\caption{Summary of the LRG and satellite galaxy populations.}
\medskip
\begin{center}
\begin{tabular} {r r r r r} 
\hline 
Galaxies & $M_{r}$ & $\sigma_{M_{r}}$ & $g-r$ & $\sigma_{(g-r)}$\\ 
\hline 
LRG & -23.3 & 0.4 & 0.94 & 0.11\\
Satellites & -21.4 & 0.2 & 0.72 & 0.05\\

\hline 
\end{tabular}
\end{center}

\medskip
A characterisation of the mean and standard deviations of the brightness and colours of galaxies in our sample. All magnitudes are k-corrected to redshift zero, and colours are quoted in the rest-frame. The satellite measurements reflect the distribution of mean satellites properties within $r_{200}$, and do not include the central LRG. 

\label{tab:brightscols}
\end{table}

\section{Halo mass estimates}\label{sec:virmass}

To assign total halo masses to objects in our catalogue, we use the
correlation between the cluster's optical richness and its mass
\citep[e.g.][]{2003ApJ...585..215Y}. This is a well-known strategy  \citep[e.g.][]{2005AdSpR..36..701V}, however, the devil is in the details: most importantly, the
choice of the suitable richness measure and the calibration of the
mass-richness relationship. In this work, we choose to link cluster's
richness to its mass via X-ray temperature. The scatter in the
richness-X-ray temperature relationship is significantly smaller
compared that of relationships between richness and X-ray luminosity or richness and velocity dispersion. 
Theoretically, X-ray temperature has also been shown to be a
robust tracer of the underlying halo mass, less sensitive to
non-gravitational effects of energy redistribution \citep[see][]{2002ApJ...576..601V}.

With this in mind, we assemble the ``anchor'' subset of SDSS clusters
with published X-ray temperature measurements. These are used to
calibrate the optical richness -- X-ray temperature relationship.
Finally, to connect richness to halo mass, we employ the X-ray
temperature-mass relationship from \citet{2006ApJ...640..691V}.  
One obvious disadvantage of our method is that only a small fraction
of our clusters have reported X-ray temperatures and therefore our
calibration sample is but a small fraction of the entire sample. However, our calibration sample covers the entire range of halo masses analysed in this paper and so we do not need to perform any extrapolation. Also, as we demonstrate below (see Section \ref{sec:sams}), there is excellent
consistency between our observed satellites profiles and those from
semi-analytic models of galaxy formation for which we use the {\it
  true} halo mass.  This agreement is non-trivial and implies that our
halo mass estimates must be quite accurate on average.

\subsection{Mass-Temperature relation in the ``anchor'' cluster sample}

The X-ray cluster temperatures in our ``anchor'' subset are mainly drawn
from three catalogues: those by \citet{2001PhDT........88H},
\citet{2008ApJS..174..117M} and the ACCEPT sample
\citep{2009ApJS..182...12C}. We also use three lower temperature samples from \citet{2003ApJS..145...39M}, \citet{2004MNRAS.350.1511O}
\citet{2009ApJ...693.1142S}. There is some overlap between the different samples used, and in cases where a cluster is reported multiple times, we assign an average X-ray temperature to the object. In these duplicate cases, the scatter between different measurements of the same object is typically within the mean 1$\sigma$ errors of the two measurements ($\sim$70 per cent within 1$\sigma$ and $\sim$90 per cent within 2$\sigma$). We find \numclusttot \ unique X-ray clusters from the above samples within SDSS DR7, within a redshift range of
0.01-0.4. These catalogues furnish us with a temperature, from which we
calculate the mass according to \citet{2006ApJ...640..691V},

\begin{equation}
\frac{M_{500}}{M_{\odot}}=\frac{M_{5}}{E\left(z\right)}\left(\frac{T}{5 \, \mathrm{keV}} \right)^\alpha\;,
\end{equation}

\noindent where $r_{500}$ is the radius which encloses the mean
density that is 500 times the critical density of the Universe,
$M_{500}$ is the mass enclosed within $r_{500}$, $T$ is the
temperature, $E\left(z\right)$ is the correction for self-similar
evolution and $\alpha=1.58 \pm 0.11$ and $M_{5}= (2.89 \pm 0.15)
\times 10^{14} \, h^{-1}M_{\odot}$ are determined by the slope of the
mass-temperature relation.  \citet{2006ApJ...640..691V} derived this
relationship from 13 low-redshift clusters (with median mass of
$M_{500} \approx 5\times10^{14} M_\odot$) under the assumption of
hydrostatic equilibrium using spatially-resolved X-ray surface
brightness and temperature profiles from {\it Chandra}.  Nagai et
al.\ (2007) have tested the methods of \citet{2006ApJ...640..691V} on
a set of clusters from cosmological simulations and showed that the
recovered masses are accurate to $\approx 15\%$ within
$r_{500}$. Recently, Sun et al.\ (2009) have shown that the
mass-temperature relation derived by \citet{2006ApJ...640..691V} also
extends down to lower-mass galaxy groups.  This was shown using a large sample of
43 groups observed with {\it Chandra} (median mass of $\approx
8\times10^{13} M_\odot$).

\subsection{Optical properties}\label{sec:optobs}

\begin{figure*}
\includegraphics[width=0.45\textwidth]{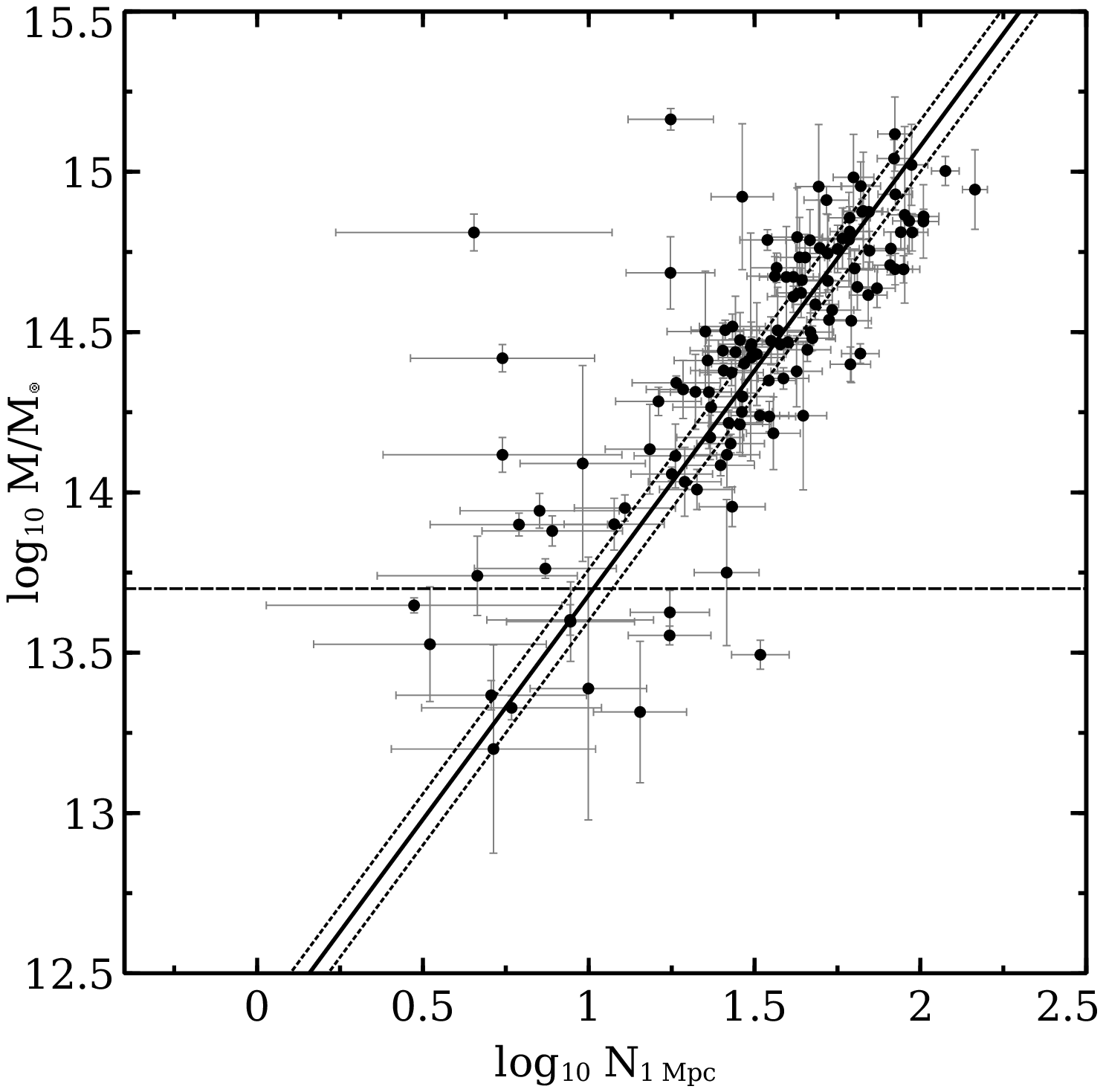}
\hspace{5mm}
\includegraphics[width=0.45\textwidth]{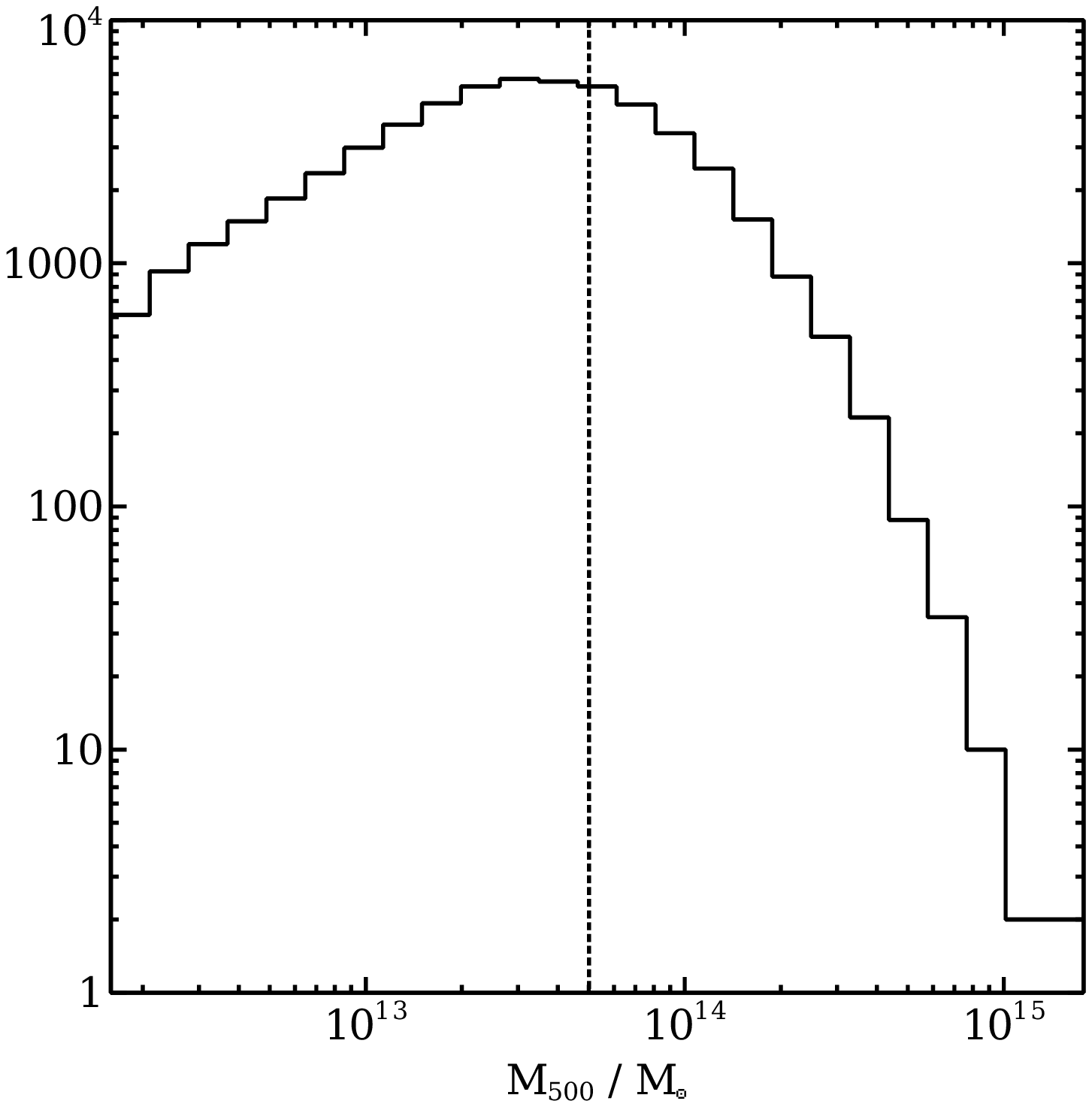}
\caption{Left: Mass within $r_{500}$ for \numclusttot
  ~``anchor'' clusters out to redshift 0.4 as a function of optical
  richness $N_{1\mathrm{Mpc}}$. The richness estimate is complete to
  the magnitude limit of $M_{r} \leq -20.5$. The solid line shows the
  best-fitting power-law which is used to evaluate $M_{500}$ for other
  clusters in our catalogue. The horizontal dashed line represents the minimum mass adopted for the sample. and the dotted lines (representing $\pm \sqrt{V}$) indicate the modelled
  intrinsic scatter in the relationship. Right: The distribution of masses in the sample obtained via the mass-richness relation shown in the
  Left panel. The sample suffers from incompleteness at the low mass
  end. The vertical dotted line shows the minimum mass adopted for the
  profile stacking analysis.}
\label{cap:xraycalib}
\end{figure*}

In the absence of any prior information about the virial radius of the
cluster, the strongest correlation between the halo mass and cluster's
environmental properties seems to arise when the galaxies are counted
in 1-2 Mpc aperture, as indicated by the analysis of the most recent
N-body simulations \citep{2011arXiv1103.0547H}. To test the
applicability of this measure to the SDSS DR7 data, we investigated
the correlation between X-ray temperature (a proxy for the total mass)
and observables such as richness and luminosity. The SDSS photometric
and astrometric data for each cluster field are obtained by running
the \texttt{q3c} radial query algorithm
\citep{2006ASPC..351..735K} on a locally available SDSS DR7 database. This retrieves all galaxies within the
five Mpc which satisfy the following magnitude and redshift
conditions:

\begin{equation}
r<21.5,\\
|z-z_{\mathrm{BCG}}| \leq 0.04(1+z_{\mathrm{BCG}}),\\
\sigma_{z}<0.2,
\label{eq:neighcond}
\end{equation}

\noindent where $r$ is the extinction-corrected $r$-band model
magnitude, $z$ is the photometric redshift of neighboring galaxies and
$\sigma_{z}$ is the corresponding photometric redshift
error. To ensure clean photometry we also exclude galaxies which satisfy the \texttt{SATURATED} photometric flag.
The apparent magnitude cut was selected to be $r=21.5$, as at this limit the SDSS completeness is close to 100 per cent and the accuracy of star galaxy separation is $\gtrsim$90 per cent. This limit was verified from repeated SDSS Stripe 82 imaging. Following \citet{2009ApJS..183..197W}, we use the variable
photometric redshift gap, which ensures most cluster member galaxies
are included regardless of cluster redshift. We first calculate the
total number of galaxies with K-corrected magnitudes $\leq -20.5$ in
the fixed 1 Mpc aperture placed on the cluster's centre (i.e., the BCG
position). To calculate the absolute magnitudes of the member galaxies, we assume that they all have redshifts equal to $z_{\mathrm{BCG}}$. Note that this \emph{absolute} magnitude limit is chosen as it corresponds
to the faintest galaxy observable (i.e. with \emph{apparent} magnitude $r<21.5$) at redshift 0.4, and ensures
completeness at lower redshifts. We then gauge the characteristic
background contribution by calculating the mean background density of
galaxies (also with $M_{r}\leq -20.5$ at the redshift of the BCG) in a series of annuli spanning a distance range of 2.5-5 Mpc
from the cluster centre.  As a check of this background density (in an annulus), we also compared with the galaxy density in randomly selected apertures, and the average difference between the density estimates was found to be independent of halo mass\footnote{The exact determination of background density is not important (provided we are internally consistent), as we are using this richness estimate purely to calibrate the halo masses.  We use randomly placed apertures for estimating the background in our analysis of satellite number density profiles in Section 4.}. The
final object richness $N_{1\mathrm{Mpc}}$ is obtained by subtracting
the estimated background contamination from the total count in the 1
Mpc cluster region. The error estimate on the richness is derived from the poisson errors on the cluster aperture and the background combined in quadrature.
The correlation between
$M_{500}$ and $N_{1\mathrm{Mpc}}$ out to a redshift of 0.4 (and a
magnitude limit of -20.5 in $r$) for our ``anchor'' sample is shown in
Fig. \ref{cap:xraycalib}.

We also investigated correlations between $M_{500}$ and other optical
observables such as cluster luminosity and richness within a scaled
aperture (designed to maximise the signal to noise of the cluster
signal above the background). However, we found that either the
scatter was increased in the case of luminosity, or, for the scaled
aperture, the improvement was too insignificant to justify the
additional complexity. In short, the simple 1
Mpc aperture proved sufficient to provide an adequate probe of halo
mass with fairly low scatter over a reasonable range of cluster
masses.

\subsection{Power-law models for mass-richness relation}\label{sec:plmodels}

We chose to parameterise the correlation between mass and richness as
follows
\begin{equation}\label{eq:masscalib}
\log \left(\frac{M_{500}}{M_{\odot}}\right) = m \log N_{1 {\rm Mpc}} + b \; ,
\end{equation}
where slope $m$ and intercept $b$ are determined using the robust
fitting algorithm of \citet{2010arXiv1008.4686H}. The method reduces
the impact of outliers, takes into account horizontal and vertical
error bars, and also modells the intrinsic scatter of the
correlation. According to \citet{2010arXiv1008.4686H}, the likelihood
function which describes the linear model is given by

\begin{equation}
\ln \mathcal{L} = K - \sum_{i}^{N} \frac{1}{2}\ln \left ( \Sigma_{i}^{2}+V \right ) -  \sum_{i}^{N} \frac{ \Delta_{i}^{2} } {2\left ( \Sigma_{i}^{2}+V 
\right )}
\end{equation}

\noindent where $K$ is a constant, $\Delta_i$ is the orthogonal
displacement of each data point from the line (Appendix
\ref{app:fit}), $\Sigma_i$ is each data points' orthogonal variance
(Appendix \ref{app:fit}), and $V$ is the intrinsic variance. This
likelihood function is maximised to give the best fit model for the
slope $m$, intercept $b$ and $V$. The outliers are pruned by rejecting
the lowest 5 per cent of contributing points to the likelihood
function on each likelihood evaluation. The fit is found to be not too
sensitive to the exact fraction of rejected points. For the ``anchor''
sample of clusters, the best-fit values are $m=1.4 \pm 0.1$, 
$b=12.3 \pm 0.1$ and $V=0.007 \pm 0.002$.

\subsection{Cluster masses}

We can use the scaling relation in Equation \ref{eq:masscalib} to
obtain $M_{500}$ estimates for all optically identified clusters which do not have direct X-ray measurements. The distribution
of the masses for all objects in our BCG-centred sample are shown in
the right panel of Figure \ref{cap:xraycalib}. 

In what follows, we adopt a lower mass limit of $10^{13.7}
\ \mathrm{M_{\odot}}$ as this is where we have only few clusters for calibration and the scatter in our mass
calibration becomes large (see left panel of
Figure~\ref{cap:xraycalib}). At this limit, the sample incompleteness becomes evident as
it coincides with the drop in the mass function of the sample as shown
in the Right panel of Figure \ref{cap:xraycalib}.  Although this mass
cut culls a large fraction of our catalogue, we are still left
with a substantial sample of \finsample \ clusters.

\subsection{The quality of the cluster catalogue}
Here we aim to check whether any significant
impurity and/or incompletness could affect the conclusions drawn
regarding the radial distribution of satellites as a function of mass.

\subsubsection{Purity}

Purity is a measure of the degree of contamination in a sample,
i.e., in our case, the number of clusters that are real compared to
spurious detections. In principle, these false detections which are caused by
interlopers in photometric redshift space should be minimised by our
imposition of a photometric redshift cut in Equation
\ref{eq:neighcond}.

We can assess the degree to which our sample suffers from impurity by
calculating `cluster' richnesses and implied masses in apertures
placed randomly on the SDSS footprint. Although there exists a small
chance that we will pick up real galaxy clusters in our randomly
placed apertures, the probability of this is low, and the averaged
recovered richness should be distributed about zero. Therefore, the
fraction of such `objects' with the mass in excess of
$M_{500}>10^{13.7}$ $M_{\odot}$ yields the {\it upper} limit to the
impurity in our sample.

The richness distribution of 1\,000 randomly positioned apertures is
plotted in left panel of Figure \ref{cap:purecomplete}, along with the
implied richness limit of 10.3. The resulting impurity fraction is
$\sim 3$ per cent. While a small fraction of these impurities will
correspond to `real' clusters, this simple calculation allows us to
place a lower limit on the sample purity for clusters with
$M_{500}>10^{13.7}$ $M_{\odot}$ of $\sim 97$ per cent. Reassuringly,
we find that this fraction does not change significantly with
redshift.

\subsubsection{Completeness}\label{sec:completeness}

The incompleteness of a sample corresponds to the number of real
clusters that are missing, i.e. in our case the number of true galaxy
clusters with $M_{500}>10^{13.7}$ $M_{\odot}$ which are lost below our
imposed mass limit due to Poisson errors. 

To assess the degree of incompleteness in our sample we generate mock 
catalogues and recover their properties using the techniques
described in Sections \ref{sec:optobs} and \ref{sec:plmodels}. A
galaxy cluster field is simulated by sampling from the projected NFW
profile \citep{1996A&A...313..697B} with a given input halo mass
($M_{500}^\mathrm{in}$), and a concentration\footnote{Our work
  (Section \ref{sec:concs}), and the work of others suggest that
  satellites are found to be a factor of $\sim 2$ \emph{less}
  concentrated compared to the parent dark matter haloes, and
  therefore we adopt a concentration of $c_{\mathrm{dm}}/2$ for the
  simulated satellites.} determined according to the relation in
\citet{2008MNRAS.390L..64D}. The total normalisation of the NFW number
density in a simulated cluster is given by:

\begin{equation}
N_{\mathrm{tot}}=N_{1\,\mathrm{Mpc}}\frac{\int_{0}^{r_{\mathrm{vir}}}f\left ( r \right )r \, dr}{\int_{0}^{1}f\left ( r \right )r\,dr}
\end{equation}

\noindent where $N_{1\,\mathrm{Mpc}}$ is given by the mass-richness
relation (Equation \ref{eq:masscalib}), $f(r)$ is the projected NFW
density profile \citep{1996A&A...313..697B}, and $r$ is the radius
from the cluster centre in Mpc. To account for Poisson-like scatter at
low cluster occupations, the number of samples generated for the given
simulated object is \~{N}$_{\mathrm{tot}}$, which is drawn from a
Poisson distribution with $\lambda=$\~{N}$_{\mathrm{tot}}$. We then
include a flat background galaxy distribution over the 5 Mpc
aperture. We find that the background density of bright galaxies
($M_{r}<-20.5$) does not vary significantly with redshift \footnote{This helps explain why the sample
  purity does not change significantly with redshift.} in SDSS DR7, and so the
background density is drawn randomly from a Gaussian with mean
$\mu_{b}=3.38$, and standard deviation $\sigma_{b}=0.79$, in units of
galaxies per Mpc$^{2}$.

\begin{figure*}
\includegraphics[width=0.45\textwidth]{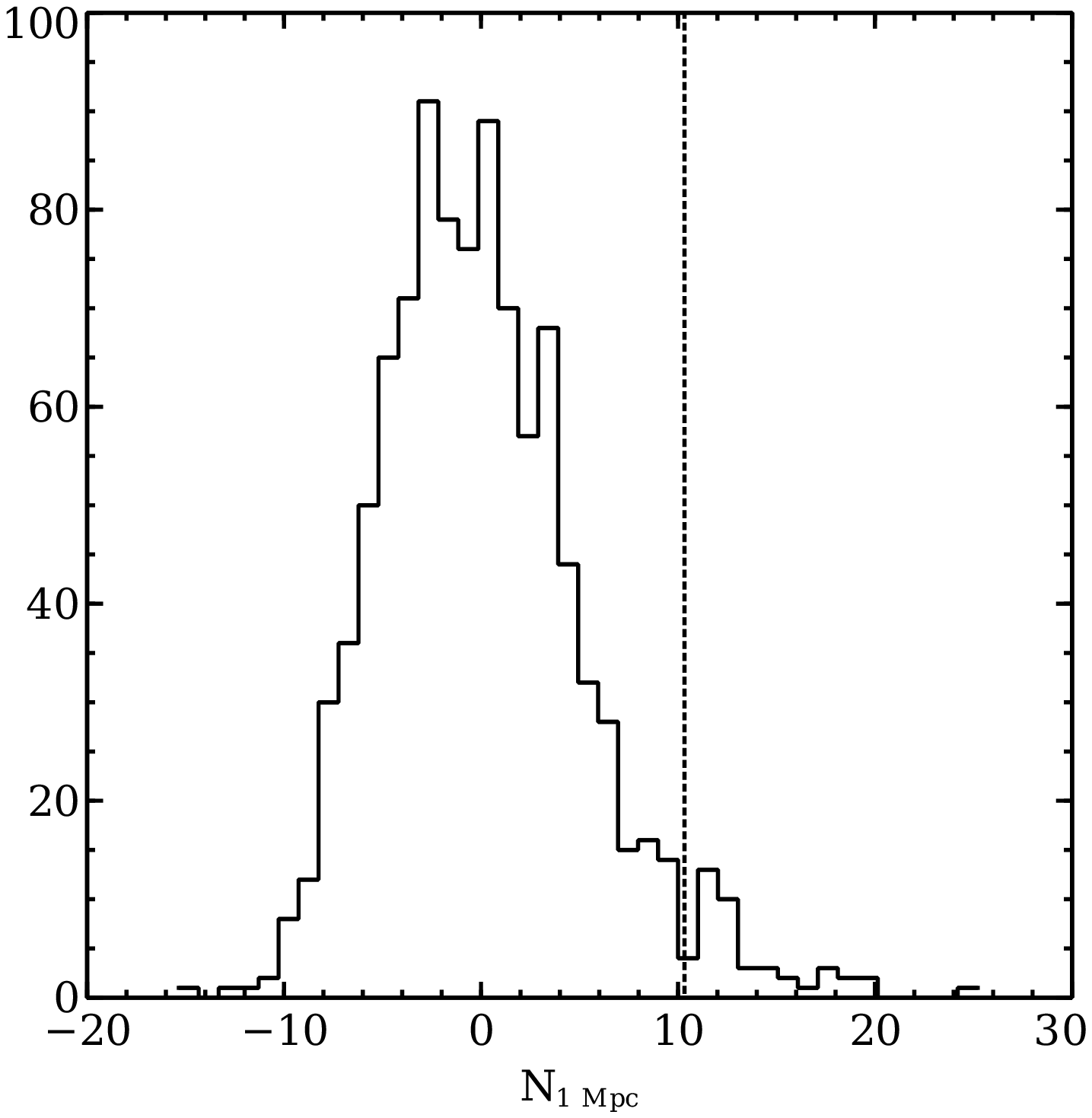}
\hspace{5mm}
\includegraphics[width=0.45\textwidth]{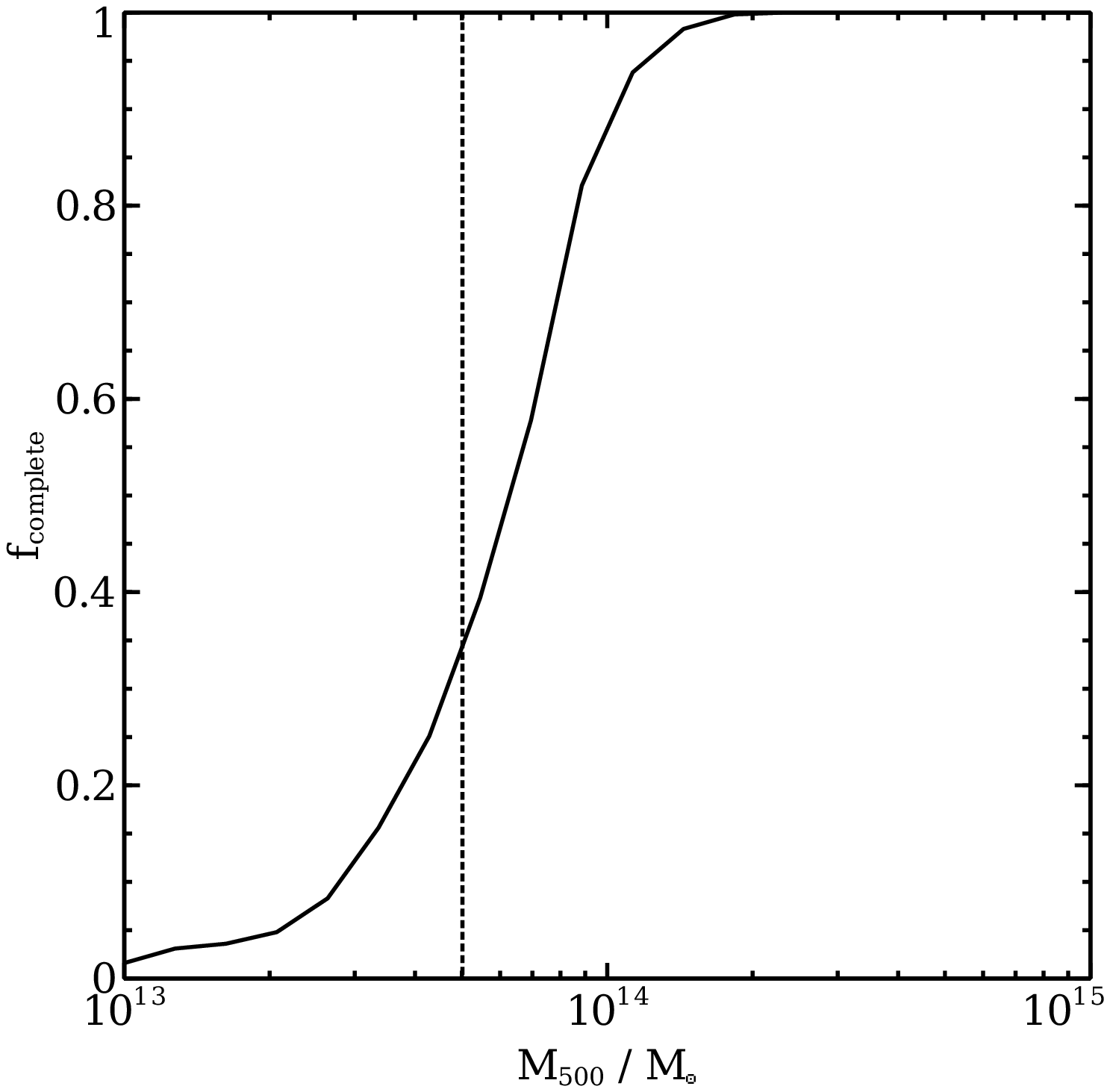}
\caption{Left: Richness $N_{1\mathrm{Mpc}}$ of 1000 fields placed
  randomly in the SDSS footprint. The fraction of objects above the
  richness limit marked by the vertical line approximates the sample's
  purity. Right: Completeness fraction as a function of halo mass for
  the cluster sample. The vertical line shows the imposed mass limit
  of $10^{13.7}$ $M_{\odot}$ corresponding to the richness cut-off in
  the Left panel.}
\label{cap:purecomplete}
\end{figure*}

A mock 5 Mpc cluster field is then analysed according to the
prescriptions in Sections \ref{sec:optobs} \& \ref{sec:plmodels} and
the implied halo mass $M_{500}^\mathrm{out}$ is obtained using
mass-richness relation. The process is repeated 1000 times to generate
a distribution of $M_{500}^\mathrm{out}$ values for a given
$M_{500}^\mathrm{in}$ value. This output distribution resembles a
Gaussian centered on $M_{500}^\mathrm{in}$. The fraction of
$M_{500}^\mathrm{out}$ values which fall below the mass limit
corresponds to the incompleteness of the sample at that
$M_{500}^\mathrm{in}$. The completeness $f_{\mathrm{complete}}$ as a
function of $M_{500}$ is shown in the right panel of Figure
\ref{cap:purecomplete}. As expected, $f_{\mathrm{complete}}$ falls to
$\sim 50$ per cent at the mass limit, and the completeness in the
lowest mass bin ($10^{13.7} < M_{500} \leq 10^{14.0}$ $M_{\odot}$) is
$\sim 70$ per cent. Reassuringly, as $M_{500}$ is increased the curve
rapidly approaches 100 per cent completeness.

A later result of this paper states that satellites in clusters are approximately \emph{half} as concentrated as the dark
matter. It is therefore important to ascertain whether the
completeness of the sample is sensitive to the concentrations of the
simulated haloes to prevent a circular argument. In our simulations,
we find that the selection function (right panel of Figure
\ref{cap:purecomplete}) is insensitive to the concentration adopted in
the range $2<c<10$. This is because the aperture used for the richness
calculation (see Section \ref{sec:optobs}) is sufficiently large to
ensure that satellites in low concentration haloes are still
predominantly located inside 1 Mpc from the cluster centre.

It is empirically known that not all massive galaxies are red.  In particular, BCGs sitting at the centers of massive clusters with very short central cooling times often show signs of active star formation and young stellar populations (e.g., \citealt{1999MNRAS.306..857C,2008MNRAS.389.1637B,2012arXiv1201.1294R}).  In these cases, the true (blue) BCG will not be designated as a BCG according to our method.  But note that as long as the cluster has {\it at least one LRG} it will be part of catalog (subject to it being overdense with respect to the background and above the adopted mass cut), even if the LRG is not the true BCG of the cluster.  To our knowledge, there is no evidence for {\it massive} groups and clusters (i.e. with at least 10 members with r $< -20.5$) with an entirely blue galaxy population (i.e., no LRGs).  Therefore we do not expect that this `blue BCG' effect has implications for our completeness calculation.  Furthermore, so long as our mass-richness relation is valid, we should be picking out the same types of groups and clusters in the observations and the models we compare to in Section 6 (even though the selection process is not identical in a procedural sense).

In the cases where the BCG is blue, the centering from our procedure (which would pick the brighest LRG) would be inaccurate.  But as we show in Section 4.2.2, the effects of inaccurate centering are minor in general and do not affect our main results or conclusions.  We can estimate the proportion of `blue-BCG' clusters which may be mis-centered due to this effect by looking at the results of \citet{1999MNRAS.306..857C}, who look at the amount of star-formation in BCGs in a modest sample of 216 ROSAT clusters. We find 149 of these clusters within SDSS, and of these, only 39 (i.e. 26 per cent) have detectable H-$\alpha$ emission.  We then cross match with the LRG sample and find an overlap of 20 with \emph{both} an LRG and H-$\alpha$ emission. Therefore, by this logic we are mis-centering $\sim$50 per cent of clusters with strong H-$\alpha$ emission in the BCG, which translates to only $\sim$13 per cent of all clusters.

\subsubsection{Comparison to other cluster catalogues}\label{sub:comments}
We can compare our cluster sample statistics with those of other
published catalogues, e.g. the MaxBCG sample
\citep{2007ApJ...660..239K}, catalogues by \citet{2009ApJS..183..197W}
(hereafter WHL) and \citet{2011ApJ...736...21S}, and the recent update
to the MaxBCG - the GMBCG sample \citep{2010ApJS..191..254H}. These
use slightly different methods of cluster identification ranging from
searching for a red-sequence in photometric data (MaxBCG), to the use
of a friends-of-friends algorithm (WHL). The purity and completeness
above the given mass limit for these published cluster catalogues are
provided in Table \ref{tab:purecomplete}.

In general, the previously published catalogues yield both purity and
completeness of $\sim 90$ per cent for clusters with
$M_{500}>10^{14.0}$ $M_{\odot}$. Our catalogue with the purity of $>
97$ per cent and completeness of $\gtrsim 95$ per cent for haloes with
$M_{500}>10^{14.2}$ $M_{\odot}$, is comparable in quality to the
previously published work.

As a final test, we repeated one of the later results in our paper
(stacking the satellite radial profiles as a function of halo mass;
Section \ref{sec:massbins}), using the MaxBCG cluster sample. The
result is included in Appendix \ref{app:maxbcg}, and is entirely
consistent with the result determined using our sample. The
consistency between the MaxBCG catalogue and our sample provides
another confirmation that it is the physics of cluster formation, and
not the selection, purity or incompleteness in optical surveys which
determines the observed radial profiles of galaxies in clusters as a
function of halo mass.

\begin{table}
\centering
\caption{Purity and completeness in other cluster samples.}
\medskip
\begin{tabular} {r r r r } 
\hline 
Survey & $M_{200}^{\mathrm{lim}}\,/\,M_{\odot}$ & Purity & Completeness\\ 
\hline 
This work & $\sim$7.5e13 & $\gtrsim$0.97 & $\sim$0.90\\
MaxBCG & 1e14 & 0.9 & 0.85\\ 
GMBCG & $\sim$1e14 & $\sim$0.9 & $\sim$0.95\\ 
Szabo et al. & 1e14 & 0.9 & 0.85\\ 
WHL & 2e14 & 0.95 & 0.9\\ 
\hline 
\end{tabular}
\label{tab:purecomplete}
\end{table}

\section{Satellites in clusters}

\subsection{Construction of the number density profiles}\label{sec:method}

For each cluster in the catalogue, galaxies satisfying the
constraints in Equation \ref{eq:neighcond} are extracted within five
Mpc of the BCG centre.  Then from a random $\pi\cdot5^{2}$ Mpc$^2$ patch
on the sky, galaxies at the redshift of the cluster are selected to
represent to the background distribution. We calculate the physical
radial distances of all neighbour galaxies with respect to the BCG
position and construct their radial distributions. This process is
repeated for all clusters in a given sub-set (e.g., halo mass bin) to yield a high
signal-to-noise number density profile. The mean background level is
obtained by co-adding random background fields to give the average
background density, which is then subtracted from each radial
bin. Finally, the stacked profiles are divided by the total number of
clusters which have contributed to the stack to give the \emph{mean}
number density profile. We provide an estimate of the error on the mean due to Poisson scatter. The true error bar will not be just Poisson distributed due to the fact that there is scatter in the shapes and normalisations of satellite profiles within the stack. This scatter is taken into account in the full modelling procedure described in Appendix \ref{app:errmodel}, and the characteristic error bar as a function of radius in the satellite profiles is calculated accordingly.

\subsection{Selection effects}\label{sec:seleffects}

\subsubsection{Bright galaxy obscuration}

As shown below the stacked satellite density profiles generally show a flattening in the
centre, at distances 20-30 kpc from the BCG (e.g. Figure
\ref{cap:stackraw}). This flattening may have a physical explanation but it could
also simply be due to the faint satellite galaxies being swamped by the
light of the bright central BCG in the SDSS imaging data.  We therefore
test the ability of the SDSS photometric pipeline
\citep{2001ASPC..238..269L} to resolve the galaxies in the wings of
the central BCG by comparing object counts in the fields that have
been observed with both SDSS and Hubble Space Telescope (HST). The
SLACS (Sloan Lens ACS Survey) galaxy-galaxy strong lensing survey
\citep{2008ApJ...682..964B} is an ideal sample for such comparison.
We use a sub-set\footnote{The entire SLACS sample consists of over 100 galaxy-galaxy lenses, but we restrict our sample to a small sub-sample observed with one orbit of ACS-WFC F814W imaging.} of 38 SLACS lenses with large BCGs with redshifts $0.1<z\lesssim0.3$. 

We obtain the ACS images from the MAST online
archive\footnote{http://archive.stsci.edu/hst/, cycle 15, proposal 10886.} and measure radial
positions, galaxy/stellar type and luminosities of possible satellites
using SExtractor software \citep{1996A&AS..117..393B}. We calibrate
the magnitudes returned by SExtractor to the SDSS magnitudes by
cross-matching positions of galaxies in the HST images and the SDSS
database, and perform the apparent magnitude cut of $r=20.35$. This cut is the apparent magnitude corresponding to the absolute magnitude limit of $-20.5$, at the mean redshift of our BCG cluster sample ($\left<z_{\mathrm{BCG}}\right>=0.29$). As well
as the loss of satellites due to the central BCG obscuration, we also
assess the potential miss-classification of stars as galaxies by the
SDSS pipeline in the following analysis.

\begin{figure*}
\includegraphics[width=0.45\textwidth]{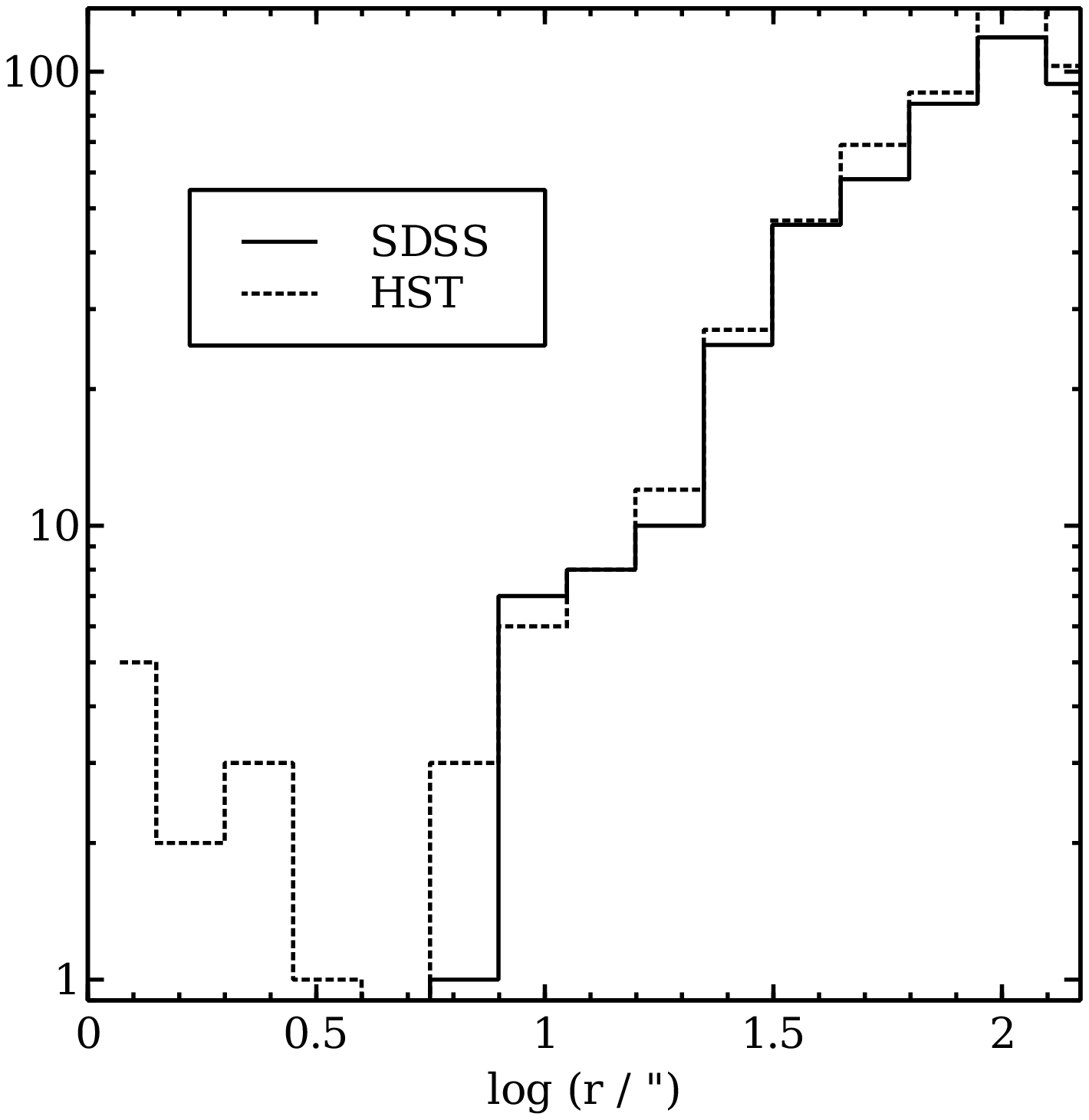}
\hspace{5mm}
\includegraphics[width=0.45\textwidth]{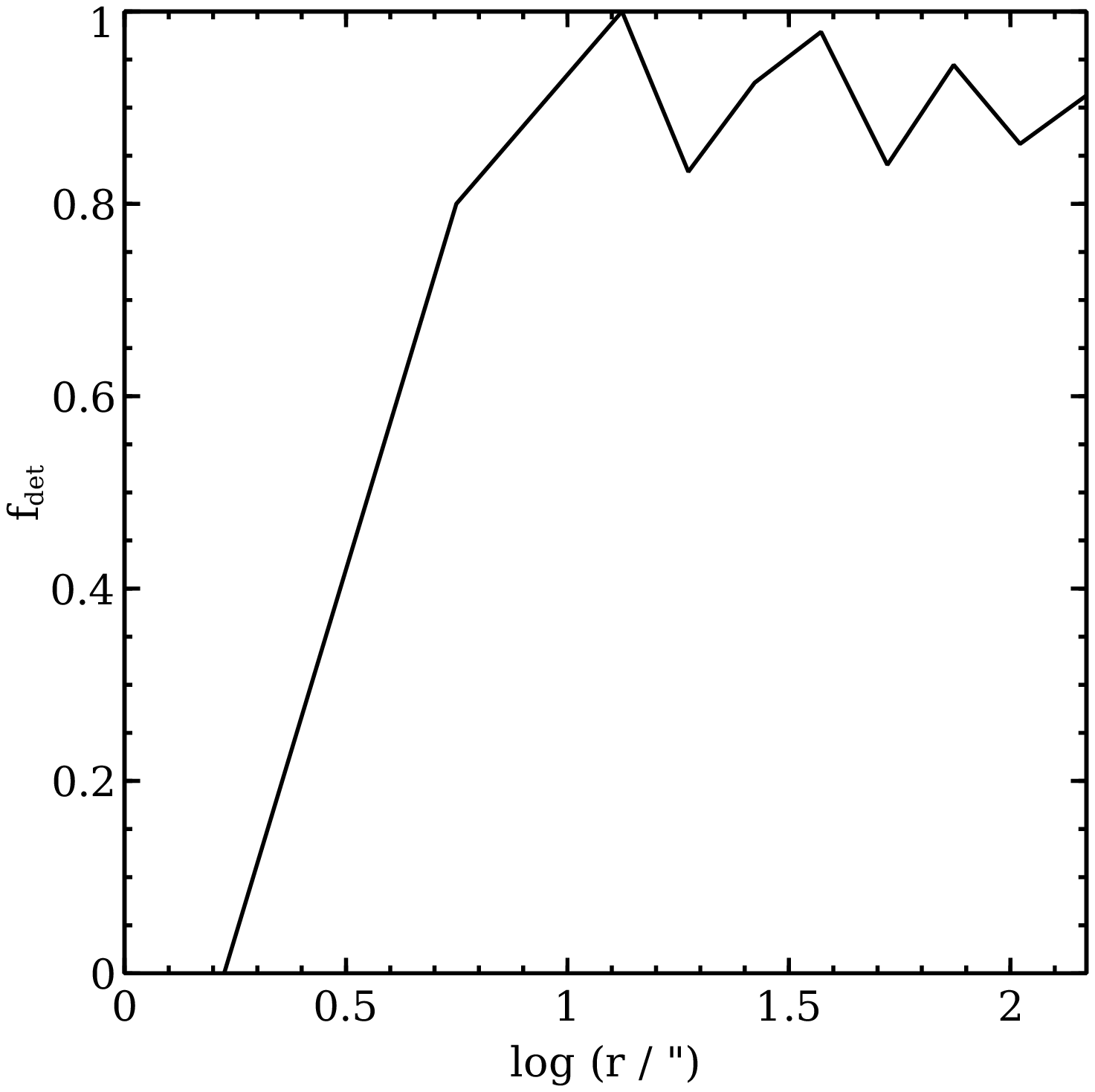}
\caption{Left: Detected number of satellite galaxies in SDSS (solid)
  and HST (dashed) for the SLACS sample imaged with the ACS. Right:
  Efficiency of detecting satellites in SDSS as a function of distance
  from the BCG.}
\label{cap:selection}
\end{figure*}

The efficiency of detecting BCG satellites in the SDSS compared to the
HST as a function of radius is given by:

\begin{equation}\label{eq:fmiss}
f_{\mathrm{det}}=\frac{\sum_i \left(C_{i,\mathrm{sdss}} \simeq C_{i,\mathrm{hst}}  
;\,t_{i,\mathrm{sdss}}\,\&\,t_{i,\mathrm{hst}}=\mathrm{galaxy}\right)}{\sum_i \left ( t_{i,\mathrm{hst}}=\mathrm{galaxy}\right )}
\end{equation}

\noindent where $C_{i,x}$ corresponds to the central position of the
satellite and $t_{i,x}$ corresponds to the star/galaxy type as
classified in the respective survey $x$. We allow for the centering
error of 1\arcsec \ when matching satellites between the two
surveys. The fraction $f_{\mathrm{det}}$ as well as the numerator and
the denominator of Equation \ref{eq:fmiss} are plotted in Figure
\ref{cap:selection}. It is clear that the SDSS detection efficiency
rapidly drops below 90 per cent when the separation from the BCG
decreases below $\sim 5$\arcsec. This number is used to define the
boundaries of the region around the BCG which is affected strongest by
incompleteness.

We have also investigated whether or not the second brightest galaxy
could obscure fainter satellite galaxies and produce non-physical
features in the radial profiles. This scenario is tested by splitting
the catalogue into faint and bright samples based on the luminosity of
the second brightest galaxy (SBG) within 200 kpc of the BCG. The
obscuration effect, if present, would show an enhanced decrement in
the radial profile of the clusters with the brighter SBG. Such a decrement
is not detected and therefore we conclude that the SBG obscuration is
not significant.

Another possible selection effect that we tested for is the degree of
galaxy-galaxy overcrowding and hence obscuration in dense clusters as
a function of redshift. The radial profiles of the clusters at high
redshifts could be systematically affected in the inner regions
as the constituent galaxies appear closer together in angular space,
and, hence, are more difficult to resolve with a modest size PSF. We
investigated this effect by calculating the covering fraction
(fraction of physical area covered by galaxy light profiles) within
200 kpc of the BCG, for the highest mass BCG clusters\footnote{The
  over-crowding of galaxies is most severe when there is dense
  clustering in the highest mass objects and so this provides an upper
  limit on the effect.} $M_{500}>10^{14.5}$ $M_{\odot}$. The galaxy light is assumed to be entirely contained within twice the petrosian radius found in the SDSS database. We found
that the covering fraction never exceeds the level of 0.01 and therefore is not a significant source of bias on our results. 


\subsubsection{BCG - cluster mis-centering}\label{sub:miscentre}

There has been considerable debate regarding the choice of the cluster
centre when computing the satellite profile, especially the effect of
possible artifacts caused by the choice of BCG as the cluster's centre
\citep{1997ApJ...478..462C, 2005ApJ...633..122H, 2005MNRAS.356.1233V}.

\begin{figure}
\includegraphics[width=\columnwidth]{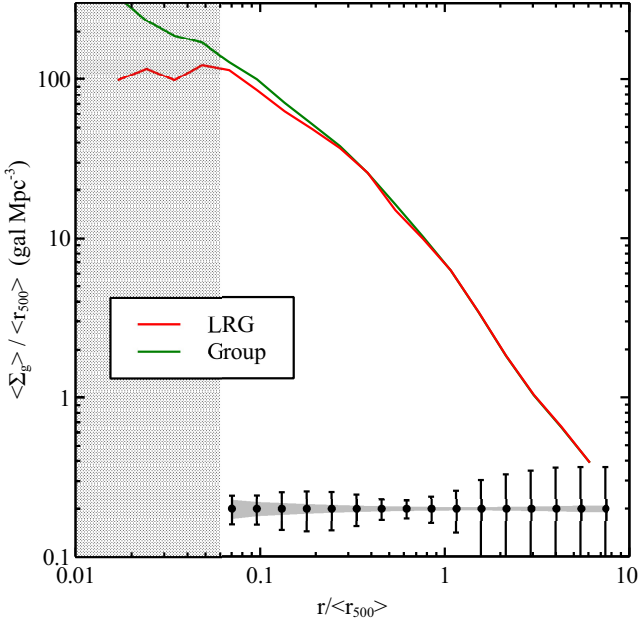}
\caption{Mean satellite number density profiles as a function of
  radius for $\sim 9\,700$ clusters centred on the BCG (green) and
  fitted centroid to the galaxy distribution (red). The black points show the representative scatter between profiles within each stack, and the shaded grey region shows the mean error bar due to Poisson scatter as a function of radius. The profiles are
  very similar beyond $\sim 0.06 \, r_{500}$, which corresponds to the
  region affected by the BCG obscuration.}
\label{cap:stackcentre}
\end{figure}

Although there are clear reasons for stacking on the BCG centre, we
have investigated the possible biases by comparing the profiles
stacked on the BCG with the profiles stacked around the fitted centre
of the satellite galaxy distribution. A tentative cluster centroid is
found by fitting a simple overdensity model consisting of constant
background and an over-density with exponential radial profile, to the
satellite galaxy field (see Budzynski et al., in preparation for
details). To ensure that the 2D density fit is reliable, we restrict
our test cluster sample to contain only significant over-densities
above the background within 300 kpc of the BCG, which corresponds to
$>$4-sigma Poisson fluctuations.  The distribution of offsets between
the BCG and the fit centre for $\sim 9\,700$ significant clusters with
is found to be a Gaussian centred around zero with the standard
deviation of 100 kpc. Importantly, the offset is not found to vary
significantly with halo mass or redshift. Figure \ref{cap:stackcentre}
shows the comparison between the cluster profiles centred on the BCG
and those centred on the fitted centroid.

Encouragingly, the profiles differ only slightly beyond the BCG
obscuration threshold of $\sim 0.06 \, r_{500}$, which indicates that
the miscentering of BCGs will not strongly affect the conclusions
drawn about the radial profile of galaxies beyond this radius.

\section{Radial profile results}\label{sec:results}

\subsection{Mass bins}\label{sec:massbins}

\begin{figure}
\includegraphics[width=\columnwidth]{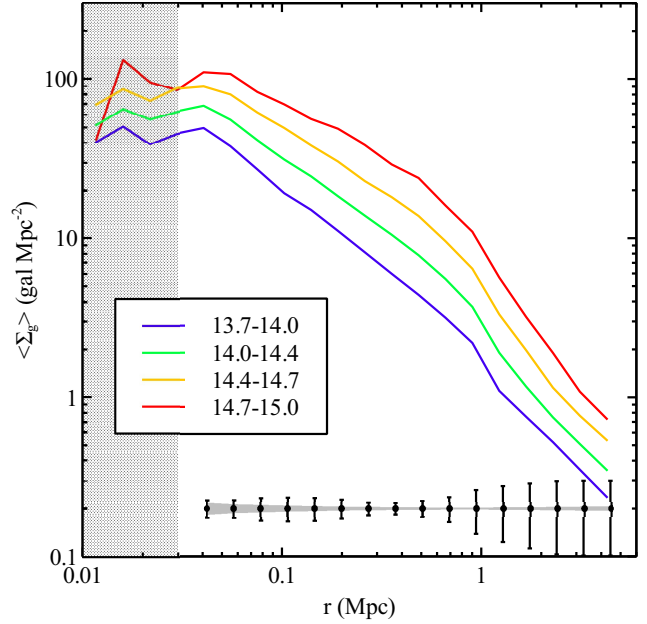}
\caption{Mean satellite number density profiles for subsets of
  clusters in bins of halo mass. The black points show the representative scatter between profiles within each stack, and the shaded grey region shows the mean error bar due to Poisson scatter as a function of radius. It is clear that the higher mass
  objects have a larger number of satellites compared to those in
  lower mass bins. The shaded (dotted) region represents the area of
  incompleteness in the number density profiles due to obscuration by
  the BCG.}
\label{cap:stackraw}
\end{figure}

In order to study the mass-dependence of the satellite profiles we have split the sample into four bins ranging from
$10^{13.7}$ to $10^{15.0} M_{\odot}$. The radial
profiles (Figure \ref{cap:stackraw}) are obtained according to the
method described in Section \ref{sec:method}. In the Figure, the
shaded (dotted) region corresponds to the region in physical space strongly
affected by the BCG obscuration effects modelled in Section
\ref{sec:seleffects}. As expected, the radial profile of the higher
mass bins have a larger normalisation than the lower mass bins. This
is due to the underlying mass-richness relation (Equation
\ref{eq:masscalib}).

The choice of the four mass bins is motivated by the degree of scatter
in the mass-richness relation (Figure \ref{cap:xraycalib}). Ideally,
one would like to divide the sample into multiple
mass bins. However, the situation is made more
complicated by the fact that the mass function of haloes is not
uniform (right panel Figure
\ref{cap:xraycalib}), and also that there is potentially significant
scatter of halo masses between bins. The choice of bin size is
therefore a trade-off between needing enough bins to probe the physics
of cluster formation which varies as a function of mass, and
having bins wide enough to not have significant mass overlap due to mass measurement errors.

\begin{figure}
\includegraphics[width=\columnwidth]{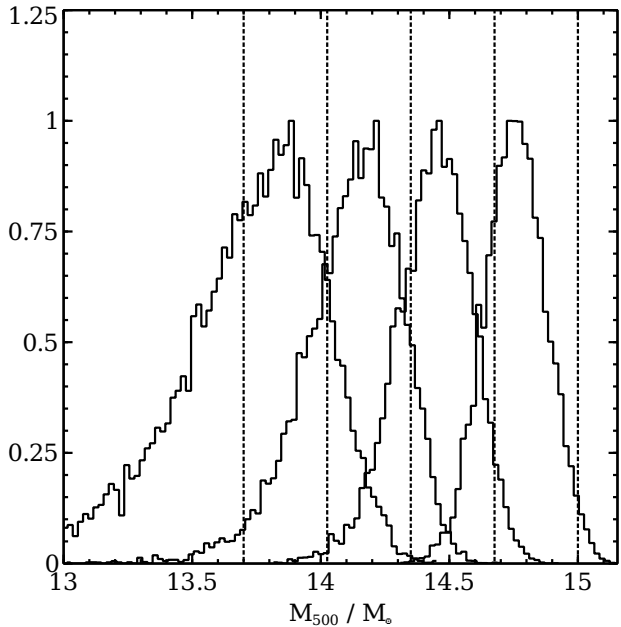}
\caption{Distributions of recovered masses for simulated clusters in
  four mass bins. The scatter becomes larger as we approach the mass
  limit of our sample ($10^{13.7} \ M_{\odot}$). The distributions in
  the higher mass bins are slightly offset to the left of the bin
  centre due to the steepness of the overall halo mass function at
  high mass end (see right panel of Figure \ref{cap:xraycalib}).}
\label{cap:massscatter}
\end{figure}

We can model the degree of scatter between the mass bins by
making use of the mock cluster simulations described in Section \ref
{sec:completeness}. The mock clusters are analysed, and mass estimates are obtained according to the prescription
in Sections \ref{sec:optobs} \& \ref{sec:plmodels}. For a given input cluster mass, the recovered mass distribution from the simulation resembles a Gaussian about the input value, with a small degree of assymmetry owing to the nature of the Poisson process. To measure the effect of scatter between bins, we first model the scatter at the centre of a mass bin and convolve the recovered distribution with a square kernel with width equal to the size of the bin. This convolution widens the initial Gaussian, which represents the fact that the true cluster mass distribution is not entirely located at the centre of the bin. A further convolution is then required to account for shape of the cluster mass function (see right panel Figure \ref{cap:xraycalib}), which is not uniform across each mass bin. The extra convolution causes the centre of the widened Gaussian to be offset to the left of the bin. This effect is more pronounced in the higher mass bins as the mass function is steeper. The resulting bin-to-bin scatter
is shown in Figure \ref{cap:massscatter}. 
The lowest mass bin is strongly affected by the inclusion of real groups and clusters with true masses below the mass limit. This effect corresponds to 50 per cent contamination for the lowest mass bin, 5 per cent for the second lowest bin and is negligible for the higher mass bins.
Despite the presence of
significant scatter between mass bins (particularly at the lowest halo
masses), distinct mass distributions within these four
bins can be
clearly seen. We can therefore be reasonably confident that any changes in
shape seen in the radial profiles as a function of halo mass are real.

It has been known for some time that the dark matter mass density
profiles look self-similar, i.e. no strong change in shape of the profile as a function of mass, \citep[e.g.][]{1997ApJ...490..493N}: if the
radial coordinate in the profile is scaled by a characteristic
radius, e.g., the virial radius, the resulting density distributions
for systems with different total mass appear nearly
identical\footnote{They are not exactly identical
  because lower mass systems are slightly more concentrated than
  higher mass systems, which reflects the fact that lower mass systems
  collapsed earlier on average at a time when the background density
  of the Universe was higher.}. It is interesting to see whether or
not such self-similarity also holds for the satellite galaxy radial
profiles. 

We produce scaled radial satellite profiles by
dividing both the distance and the number density by $r_{500}$ (the latter is required since $\Sigma_g$ is a {\it surface} number density therefore we must take into account that more massive objects have larger line-of-sight lengths in physical units).
The stacked profiles are normalized by the mean value of $r_{500}$ for the four mass bins we consider.  
The mean value of $r_{500}$ is calculated as

\begin{equation}\label{eq:r500}
\left<r_{500}\right>=\left( \frac{3}{2000 \pi \rho_{c}\left(z \right)} \left<M_{500}\right>  \right)^{1/3} \;,
\end{equation}

\noindent where $\rho_{c}\left(z \right)$ is the critical density of
the universe at the redshift of the cluster, and
$\left<M_{500}\right>$ is the mean mass in a given mass bin.

\begin{figure*}
\includegraphics[width=0.45\textwidth]{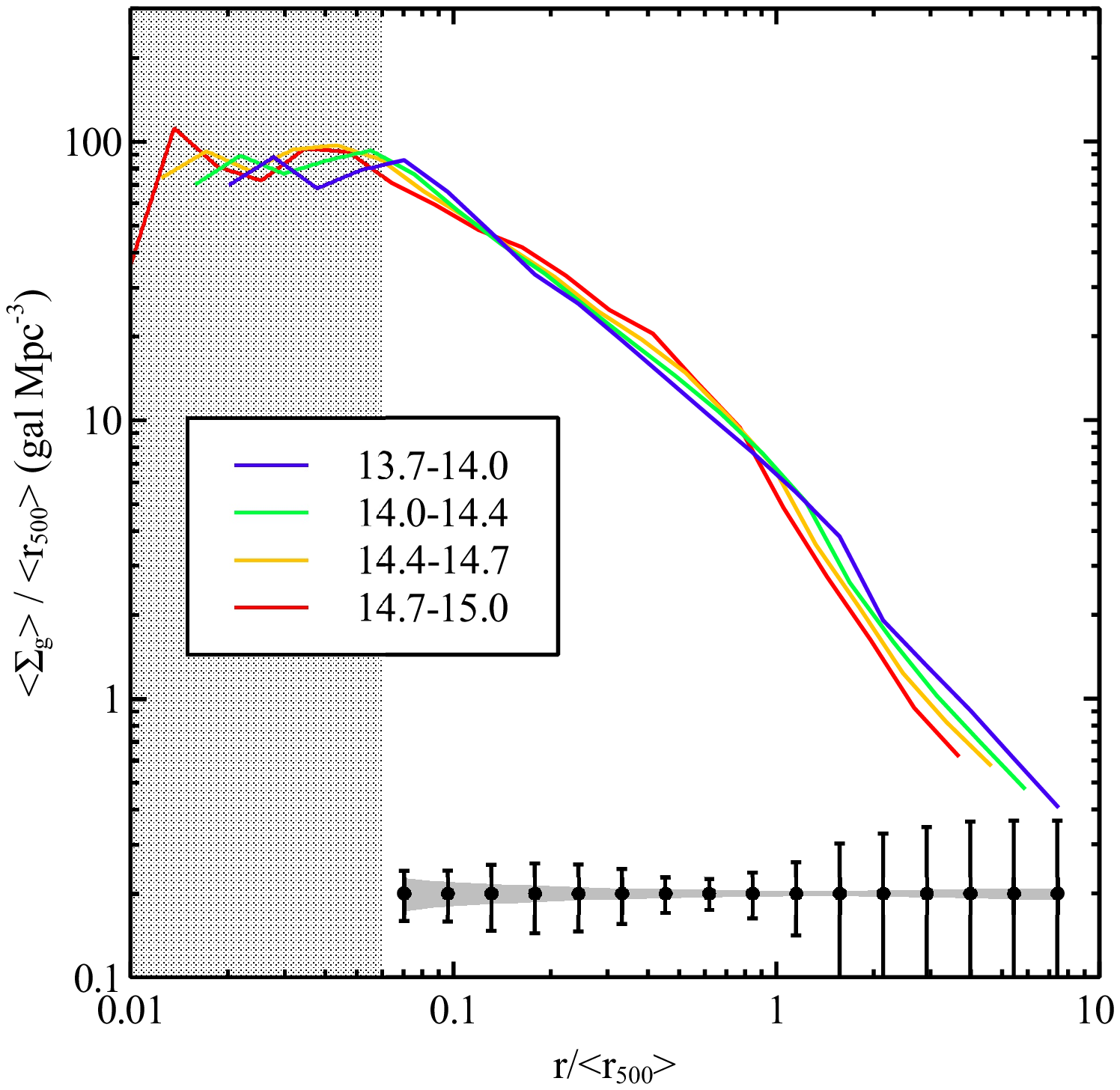}
\hspace{5mm}
\includegraphics[width=0.45\textwidth]{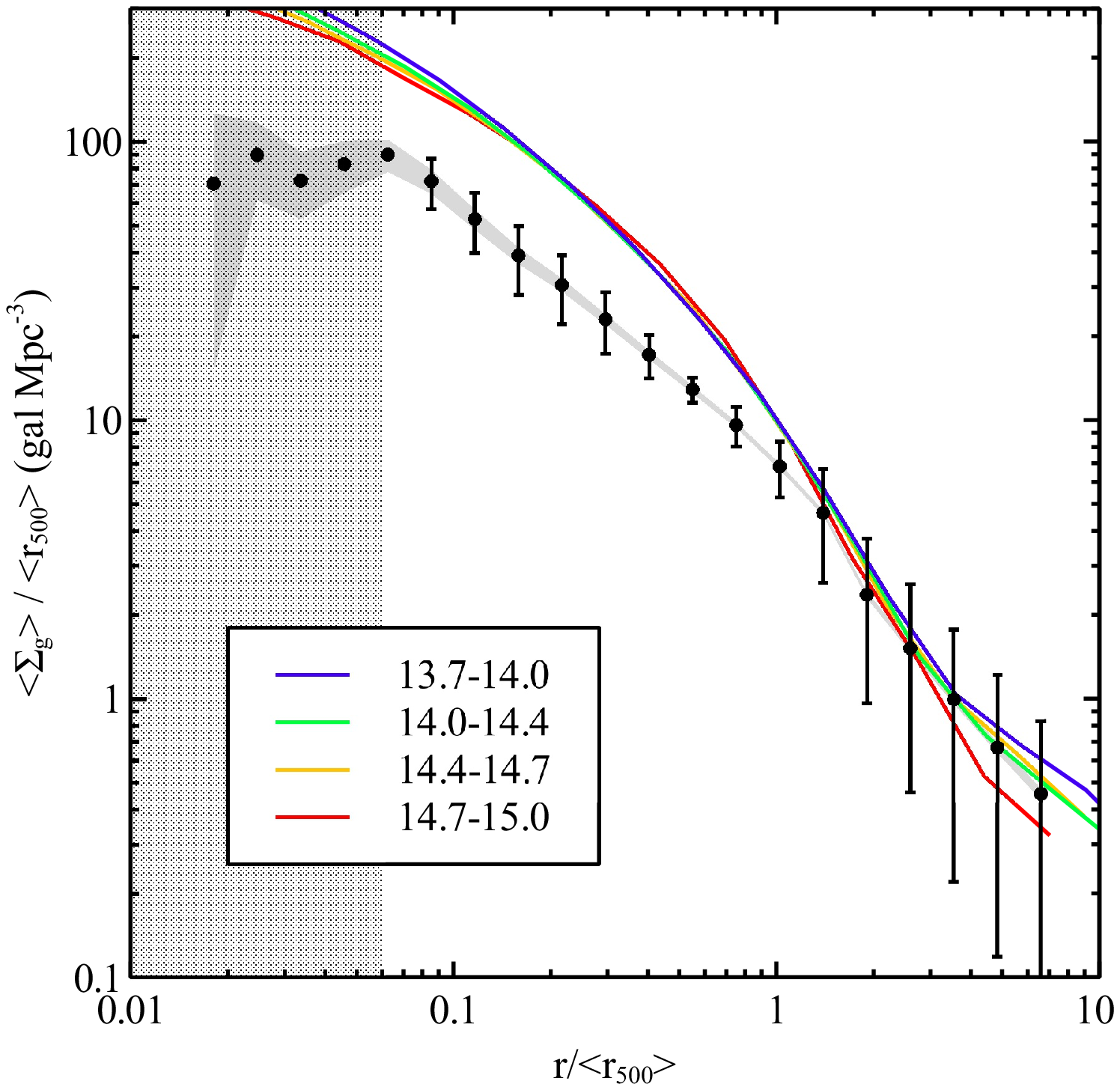}

\caption{Comparison of the observed scaled profiles with the simulated
  ones. Left: Mean satellite number density profiles split into bins
  of halo mass. Both axes have been scaled by the mean $r_{500}$ in a
  given bin to remove the mass dependence.  The black points show the representative scatter between profiles within each stack, and the shaded grey region shows the mean error bar due to Poisson scatter as a function of radius. The shaded (dotted) region represents the area of
  incompleteness in the number density profiles due to obscuration by
  the BCG. Right: Observed satellite
  number density profiles (solid black dots) in a single mass bin (from $10^{13.7}-10^{15.0}$ $M_{\odot}$) compared to the dark matter profiles of the
  Millenium clusters in the mass bins shown in the left panel (colour
  curves). The normalisation of the dark matter profiles have been multiplied by an arbitrary (constant) factor.}
\label{cap:stacksimcomp}
\end{figure*}

Figure \ref{cap:stacksimcomp} shows these scaled satellite profiles for
the four mass bins considered.  Remarkably, this scaling procedure
removes most of the mass dependence in the satellite density
distributions.  Only a small residual difference remains in that the
radial profiles of lower mass haloes appear to be slightly more peaked
compared to those of the high mass clusters.

\begin{figure}
\includegraphics[width=\columnwidth]{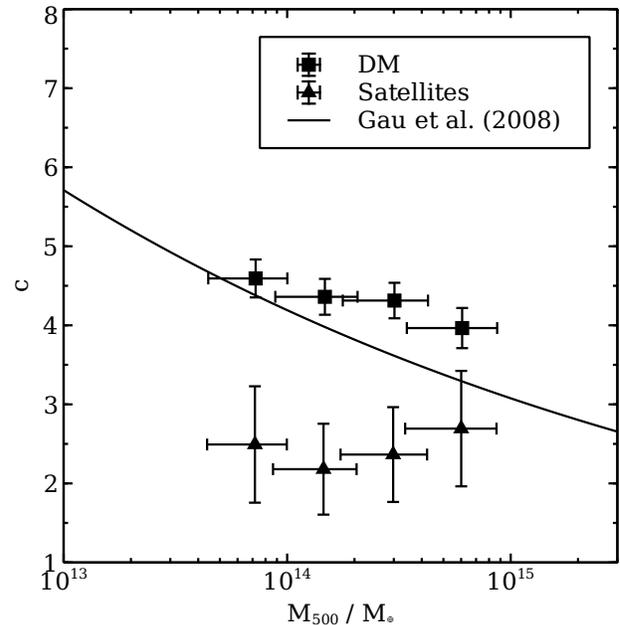}
\caption{Concentration as a function of mass
  for the observed satellite number density profiles (triangles) and
  the Millenium dark matter clusters (squares) shown in Figure
  \ref{cap:stacksimcomp}. The solid line corresponds to the predicted
  mass-concentration relations from \citet{2008MNRAS.387..536G}.}
\label{cap:massconc}
\end{figure}

\subsection{Satellite concentration}\label{sec:concs}

How well do satellites trace the underlying dark matter distribution?
We can test whether the satellite behaviour is matched by the overall
dark matter halo density distribution by looking at the projected dark
matter density profiles from the Millenium Simulation
\citep{2005Natur.435..629S}. We select haloes in the same mass bins as
the observations and create density profiles as a function of
$r/r_{500}$.  The DM simulation profiles are arbitrarily normalised by multiplying them by a constant factor so that the profile corresponding to the [14.0-14.4] halo mass bin passes through the observed galaxy profile at $3 r_{500}$ (right panel of Fig. \ref{cap:stacksimcomp}).  
%
%
As expected, the dark
matter profiles are nearly self-similar.  Interestingly, the dark
matter profiles exhibit the same shape dependence as a function of
mass as the satellite profiles (i.e. slightly more peaked at lower
mass). This dark matter shape change is well documented in terms of
the mass-concentration relationship in N-body simulations
\citep{1997ApJ...490..493N, 2001MNRAS.321..559B, 2008MNRAS.390L..64D, 2008MNRAS.387..536G},
in which the lower mass haloes appear more concentrated compared to
the high mass ones. This concentration dependence is thought to be due
to the fact that more massive haloes collapse at a later cosmic time,
where the background density of the universe is lower.  It is not immediately
obvious, though, that the satellites should preserve this trend, since
presumably a large fraction of the satellites we see in orbit today
were accreted not that long ago.  The fact that the overall shapes of
the satellite profiles are systematically different from the dark
matter but that both are individually nearly self-similar is
therefore quite intriguing.

We can quantify the difference in the shapes of the dark matter and
the satellite density distributions by measuring their concentrations
as described by the NFW profile (Figure \ref{cap:massconc}). We fit
the projected NFW profile \citep{1996A&A...313..697B} to the satellite
density profiles over the radial region $0.07<r/r_{500}\leq1.5$. Figure \ref{cap:massconc} shows the
obvious drop in halo concentration as a function of mass. The best-fit power-law relation
of \citet{2008MNRAS.387..536G}, which was derived from fitting over a much wider range of halo masses, is over-plotted for
comparison\footnote{The small discrepancy
  between our derived mass-concentration relation for the simulated clusters and the best-fit power-law relation of \citet{2008MNRAS.387..536G} may be the result of those authors fitting over a wider range of halo masses.  Another difference is that our concentrations are derived by fitting to the projected surface mass density profiles, rather than to 3D density profiles.}.
One of the most striking observations is that the concentration of satellites is roughly a factor
of two smaller than the dark matter across the full range of halo
masses we have explored.

This systematic difference between the projected DM and galaxy radial distributions
is generally consistent with the findings of previous studies. For example,
\citet{2004ApJ...610..745L} in their study of 2MASS clusters in the
K-band, find that the distribution of satellites has the mean
concentration of 2.9. Additionally, two separate analyses of the CNOC
clusters by \citet{1997ApJ...478..462C} and
\citet{2000AJ....119.2038V} find the concentrations of 3.7 and 4.2
respectively. Although our concentrations are roughly a factor of 1.5
lower than these studies, we find a general qualitative agreement in
that we see a relatively shallow dependence of the concentration on
the halo mass. This shallow dependence is apparently discordant with
the findings of \citet{2005ApJ...633..122H} and
\citet{2006ApJ...647...86C}, who find that the concentration is a very
strong function of optical richness.  It is likely that this
difference can be attributed to the differing characteristic radii used to scale the radial coordinates of the profiles (see the discussion in Section 1).  Namely, \citet{2005ApJ...633..122H} define their characteristic radii ($R_{200}^{N}$) with respect to the background density of galaxies, whereas our characteristic radii are defined with respect to the critical density of the Universe from X-ray measurements (see Equation \ref{eq:r500}).
Indeed, using clusters in common between our catalog and that of the MaxBCG catalog, we find a scaling of $r_{200} \approx 0.65 r_{500} \sim 1.4 R_{200}^{N}$, which accounts for the steep richness dependence of their derived concentrations.

Finally, a possible explanation for the offset between dark matter and
satellite concentration is the fact that the galactic subhaloes are
strongly affected by tidal evolution and merging process within the
inner regions of the cluster
\citep{2005ApJ...618..557N,2006ApJ...647...86C}. We will investigate the physics of subhalo
disruption and merging in the context of semi-analytic galaxy
formation models in Section \ref{sec:sams}.


\subsection{Satellite profile shape as a function of other properties}

Since the radial distribution of satellite galaxies does not vary strongly with halo mass, we can combine all galaxies into one cluster mass bin, ranging from $10^{13.7}$ to $10^{15.0} M_{\odot}$, and investigate the dependence of the profile (particularly its shape) on other properties.  We now investigate whether the stacked radial profiles of satellite
galaxies depend on redshift, BCG luminosity (or BCG luminosity fraction), and the luminosity and colour of the satellites themselves.

\subsubsection{Redshift}\label{sub:redshiftdep}
Splitting the sample according to cluster redshift reveals a small change in the shape of the derived radial distribution (Fig. \ref{cap:stackscaled} \emph{(a)}).  In particular, within $r \approx 0.2 r_{200}$, the profile is slightly shallower in our lowest redshift bin with respect to that of the intermediate and high redshift bins.  Beyond this radius, the shape and normalisation of the profiles in all three redshift bins are consistent with no evolution.

For reference, cosmological N-body simulations predict that the concentration should {\it increase} with decreasing redshift for a cluster of fixed mass (see, e.g., \citealt{2008MNRAS.390L..64D}).  However, over the relatively narrow range of redshift that we probe here, the DM concentration for a cluster of fixed mass would only vary by $\approx 7.5\%$ according to \citet{2008MNRAS.390L..64D}.  In any case, it is not immediately obvious that the shape of the galaxy and DM radial distributions should evolve by the same amount or even in the same sense.  Indeed, dynamical friction and tidal stripping/disruption could plausibly act to reduce the concentration of the satellite distribution (for clusters of fixed mass) over cosmic time.

\subsubsection{BCG luminosity}

\begin{figure*}
(a)
\includegraphics[width=0.40\textwidth]{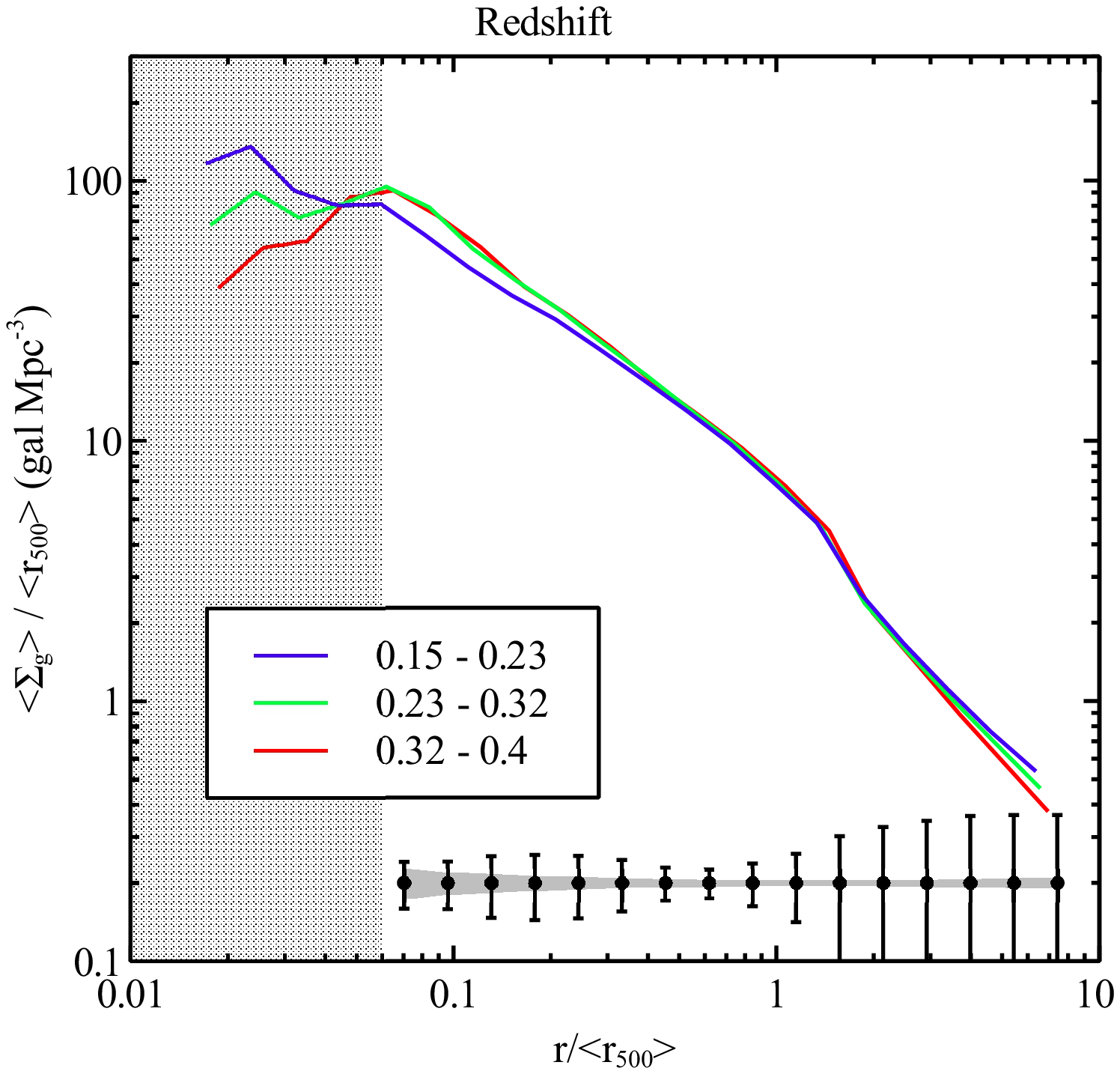}
\hspace{5mm}
(b)
\includegraphics[width=0.40\textwidth]{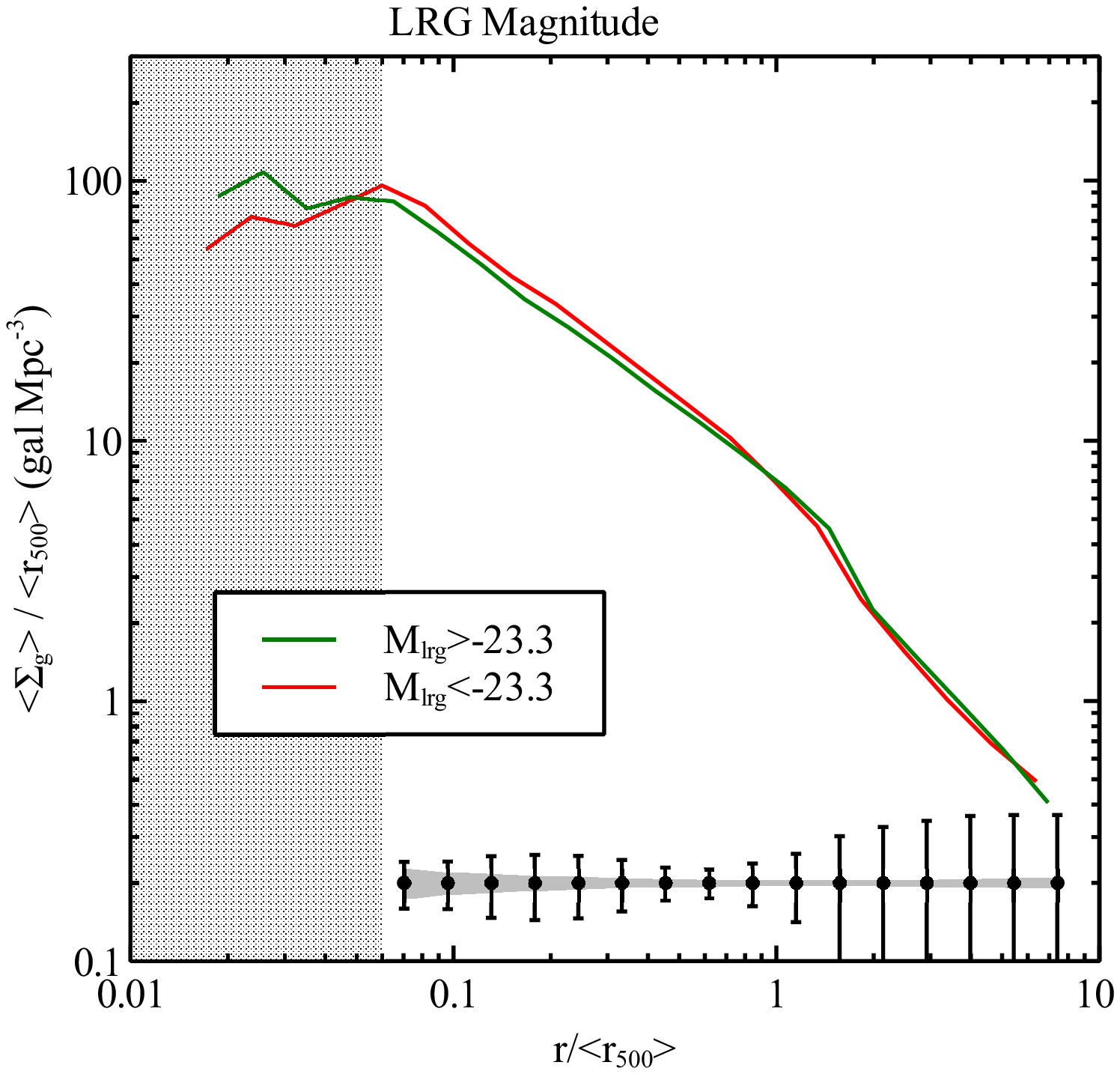}
\vspace{5mm}

(c)
\includegraphics[width=0.40\textwidth]{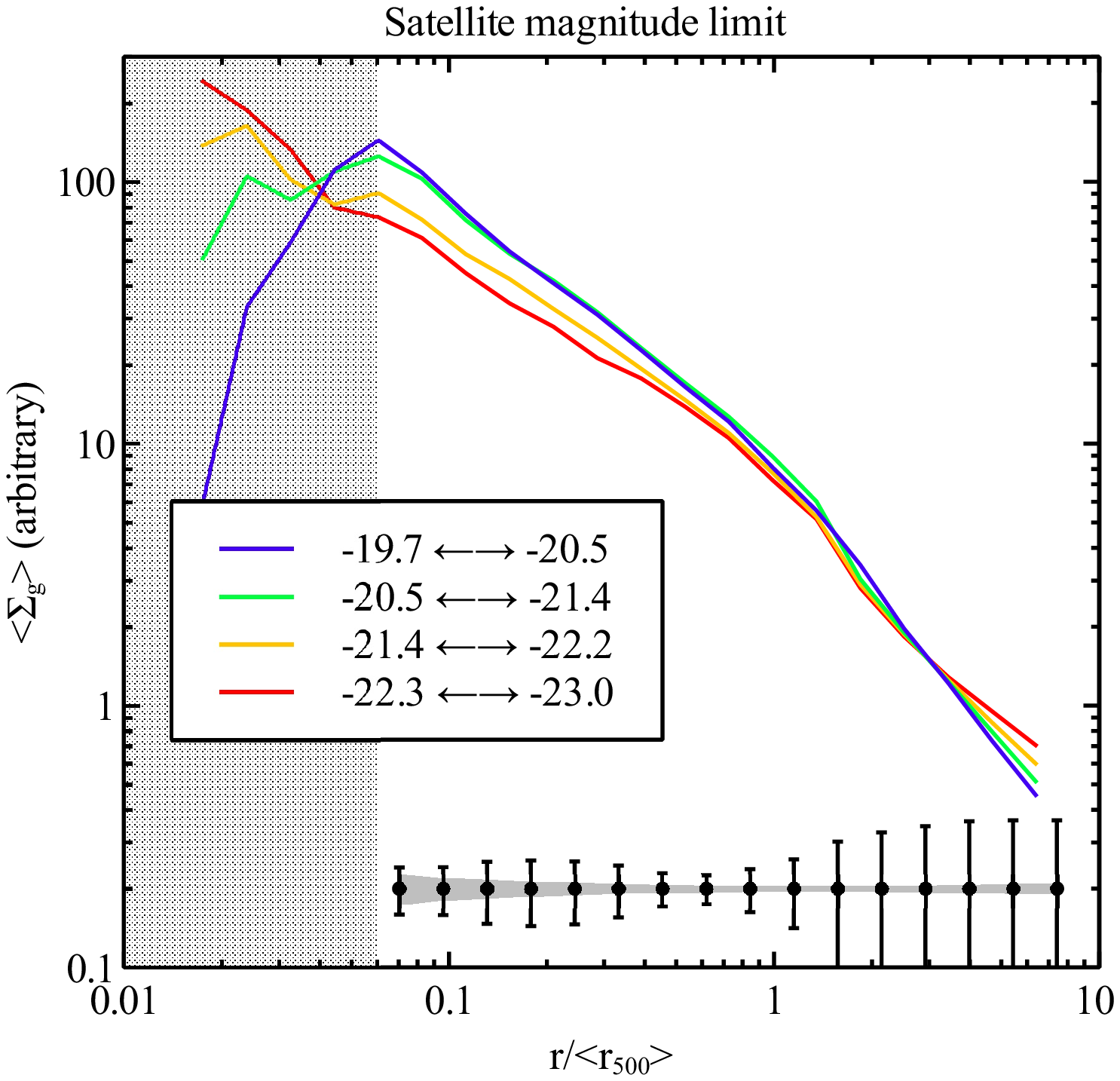}
\hspace{5mm}
(d)
\includegraphics[width=0.40\textwidth]{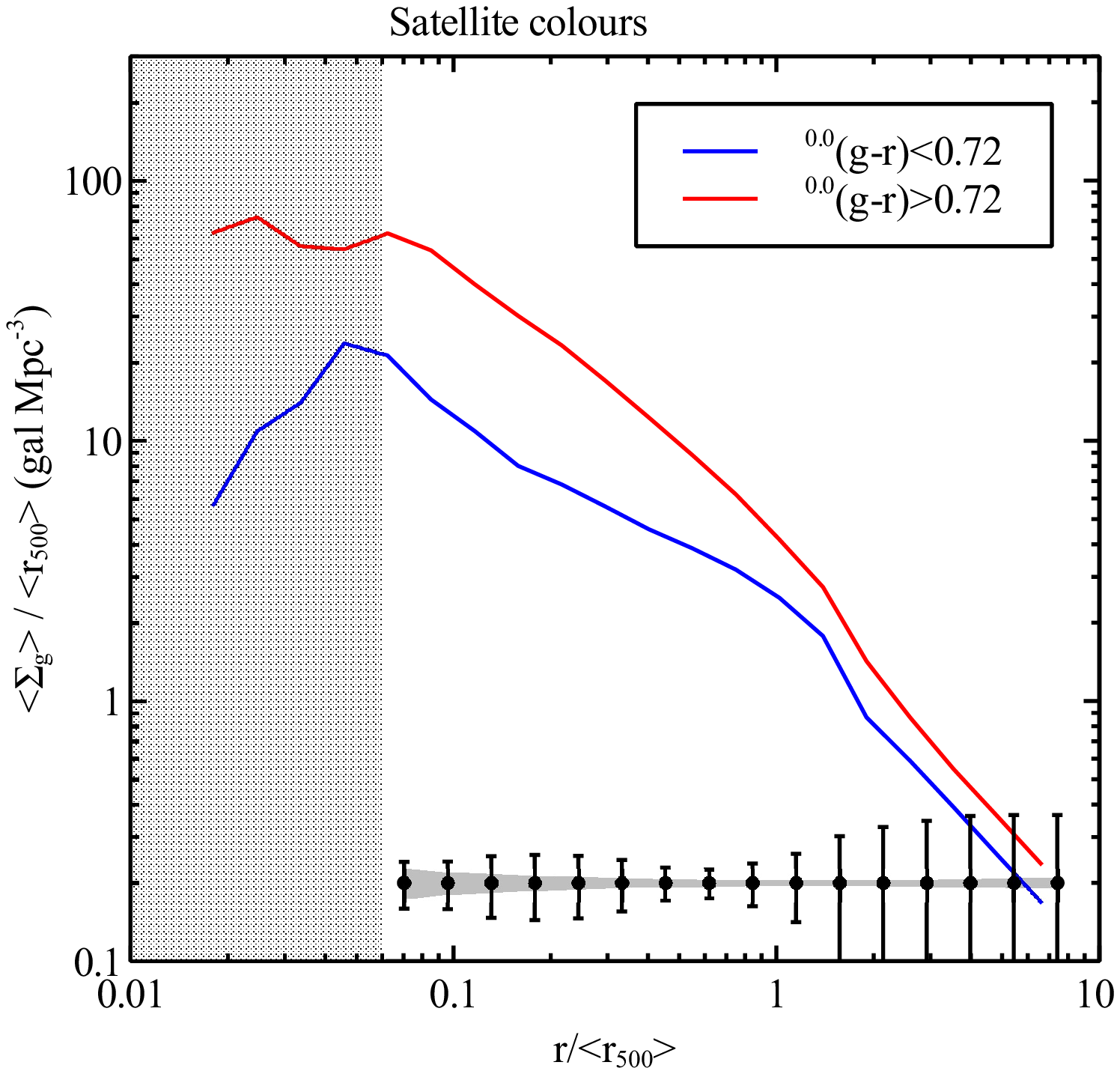}
\vspace{5mm}

\caption{Mean satellite number density profiles split into bins of
  cluster redshift, BCG luminosity, BCG luminosity fraction and
  satellite magnitude limit. The black points show the representative scatter between profiles within each stack, and the shaded grey region shows the mean error bar due to Poisson scatter as a function of radius. All horizontal axes have been scaled by the mean $r_{500}$ in a given bin to remove the
  mass dependence. The shaded (dotted) regions represent areas of
  incompleteness due to BCG obscuration. Top Left (a): Cluster sample
  is split by redshift. Top Right (b): The sample is split according
  to the mean BCG magnitude. Bottom Left (c): The sample
    (limited to $z<0.3$) split by absolute magnitude limit of the
    satellites. There profiles have been arbitrarily normalised
    further out to $\sim 3r_{200}$. Bottom Right (d): Sample split by satellite colour (i.e. $g - r$ brighter and fainter that 0.72).}
\label{cap:stackscaled}
\end{figure*}

Splitting the sample according to relatively faint/bright BCGs (about the median BCG magnitude of $M_r=-23.3$),
reveals a small (but statistically significant within the Poisson error bars) difference between the two samples (panel (b) of Figure
\ref{cap:stackscaled}).  Naively speaking, we might expect to see a central decrement in the satellite profiles for clusters with brighter than average BCGs (assuming the BCG is built from cannabalizing satellites).  However, the observed trend actually goes the other way.  A more mundane explanation for the trend is that there is a known dependence (albeit a relatively weak one) on the luminosity of the BCG and total cluster mass.  When we split the sample based on BCG fraction (ratio of
BCG luminosity to the background-subtracted cluster luminosity within 1 Mpc), the shape dependence of the satellite profiles all but disappears.  We therefore find that when the halo mass dependence of the BCG luminosity is taken out, there is no obvious dependence of the satellite radial distribution on BCG luminosity.

\subsubsection{Satellite luminosities}

We turn our attention now to the brightness of the satellites
themselves and investigate the dependence of the shape of the profile
in four satellite luminosity bins over the range $-20.5 > M_{r} \geq
-24.0$ (panel (c) of Figure \ref{cap:stackscaled}). We find that the
concentration of satellites falls slightly as their brightness
increases. This is opposite to the finding of
\citet{2004ApJ...617..879L} and \citet{2006ApJ...647...86C} who find
that the fainter satellites show a small dip in the central regions
relative to the bright satellites. However, these studies only find a
very small difference between faint/bright satellites and also stack a
far fewer number of clusters than the number we present in this
study.

Our results imply that the dwarf-to-giant ratio (DGR - the
ratio of faint-to-bright galaxies), increases into the centre of the
halo. This conclusion is in qualitative agreement with the work
of \citet{2000ApJ...539..136Z}, who find that the DGR decreases with
halocentric radius albeit in a sample of lower masses
and smaller redshifts than probed in this
work. \citet{2000ApJ...539..136Z} propose a number of scenarios to
explain the behaviour of the DGR with radius, particularly that
brighter and more massive galaxies are subject to larger amounts of
dynamical friction and thus more frequently merge with the BCG than
their fainter counterparts. This effect leads to a deficiency of
bright, massive galaxies in the centre of the halo. Our results would
therefore suggest that such a mechanism may occur in the high mass end
of clusters as well as in the poorer groups looked at by
\citet{2000ApJ...539..136Z}.

\subsubsection{Satellite colours}

A split of the satellite profile stacks into red and blue satellites
(done about the median colour\footnote{k-corrected to redshift zero.}
of $g-r\sim0.72$) yields a striking difference between their profile
shapes. An impressive deficit\footnote{It is worth noting that the
  slight rise at $\sim 0.1\,r_{500}$ in the profile of blue galaxies
  could be an artifact possibly caused by the differential effects of
  BCG obscuration for the red/blue galaxies.} of blue satellites in
the centre of the halo relative to red satellites is consistent with
what is found by \citet{2005MNRAS.361..415C}, who observe that the
profiles of red galaxies ($c=3.9$) are much more concentrated than
those of blue galaxies ($c=1.3$) in a large sample of 2PIGG
groups. This conclusion is echoed by \citet{2006MNRAS.366....2W,
  2009MNRAS.394.1213W} who find that the
early-type\footnote{`Early/late-type' is a definition based on the
  star-formation rate of the galaxy, but observationally these are
  very strongly correlated with red/blue colour of galaxies.}
fraction of galaxies decreases with increasing halocentric radius, and
vice-versa for late-types.

A couple of mechanisms commonly invoked to explain the red/blue
behaviour are ram-pressure stripping and strangulation. Ram-pressure
stripping \citep{1972ApJ...176....1G} causes rapid removal of cold
gas, and strangulation \citep{2000ApJ...540..113B} occurs when
satellites are deprived of their hot gas reservoir.  Both processes
will result in a supression of star formation in the satellite galaxies and
can transform late-type galaxies (blue) into early-types (red). These
processes which quench star-formation become more severe as we move
towards the centre of the halo, which explains the large observed
discrepancy between red and blue galaxies within $r_{500}$.

\begin{figure*}
(a)
\includegraphics[width=0.40\textwidth]{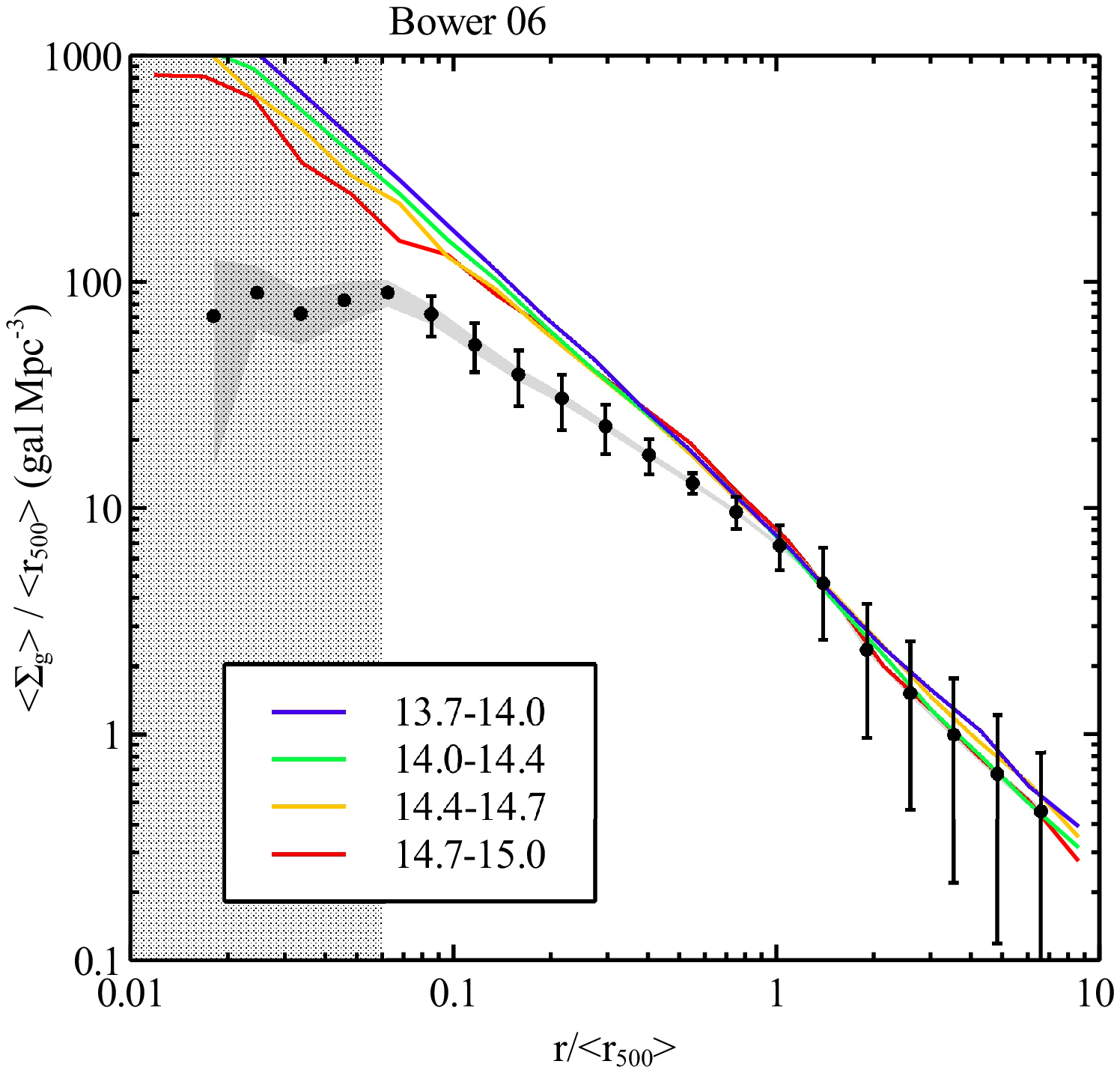}
\hspace{5mm}
(b)
\includegraphics[width=0.40\textwidth]{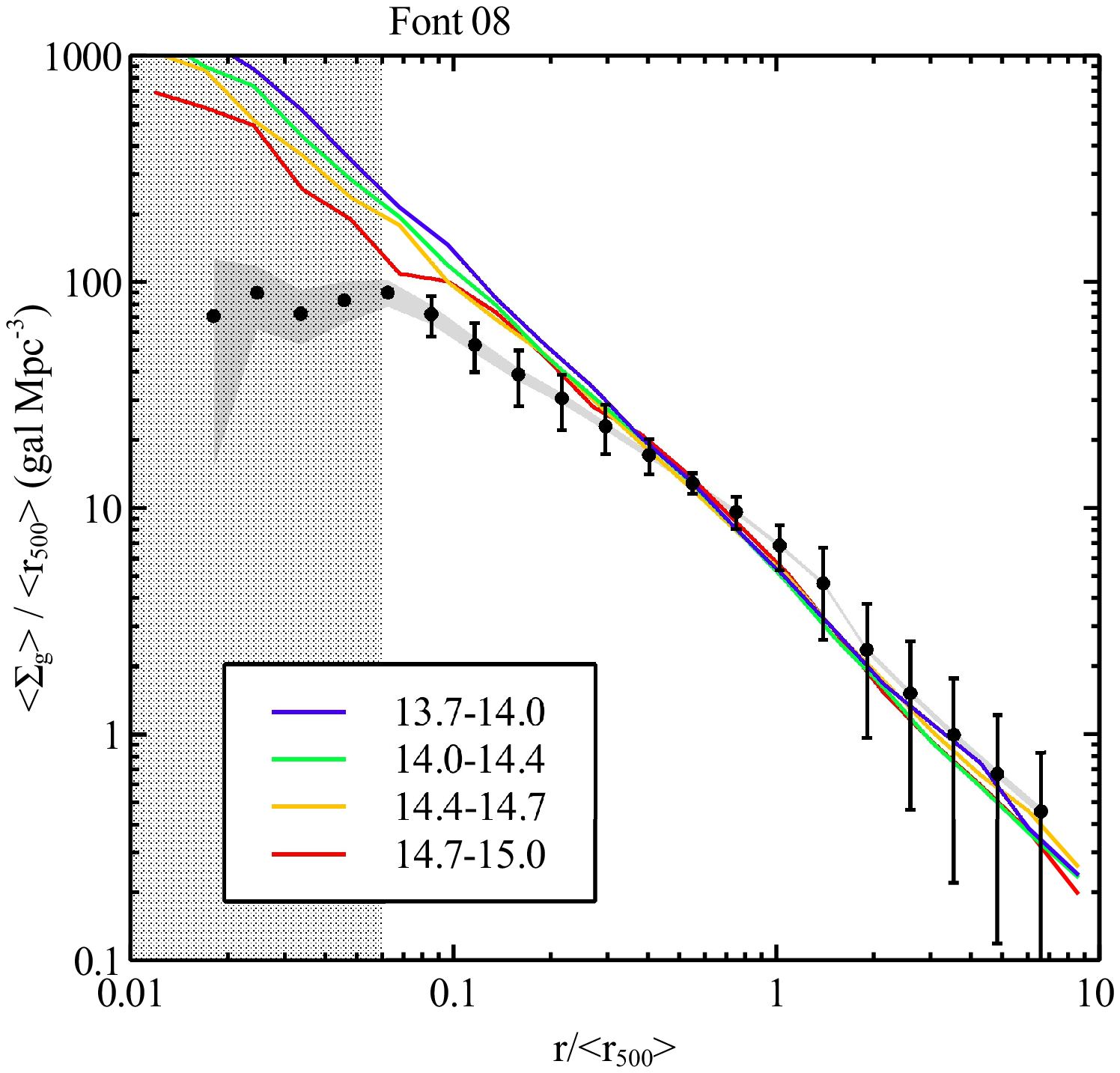}
\vspace{5mm}

(c)
\includegraphics[width=0.40\textwidth]{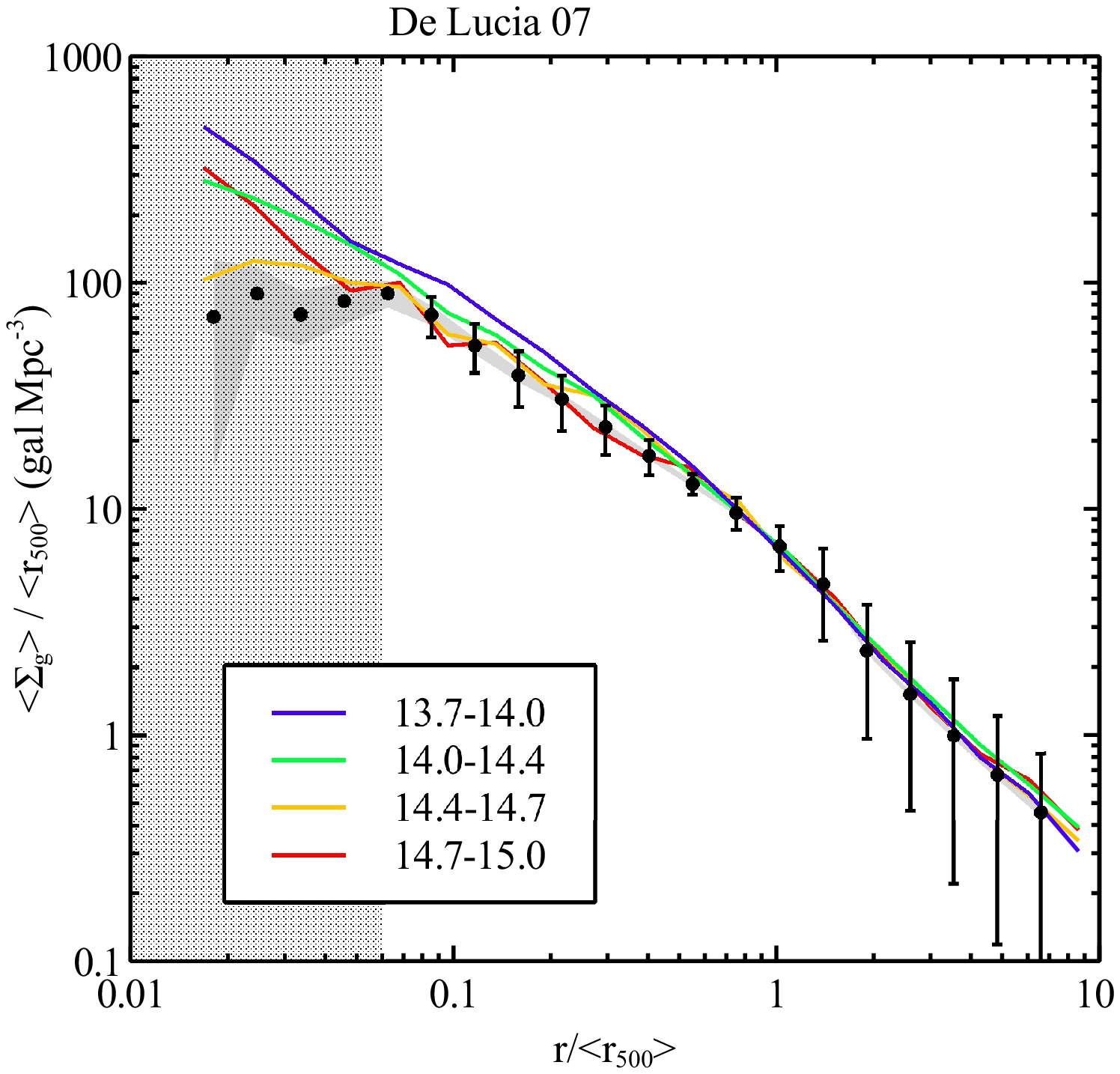}
\hspace{5mm}
(d)
\includegraphics[width=0.40\textwidth]{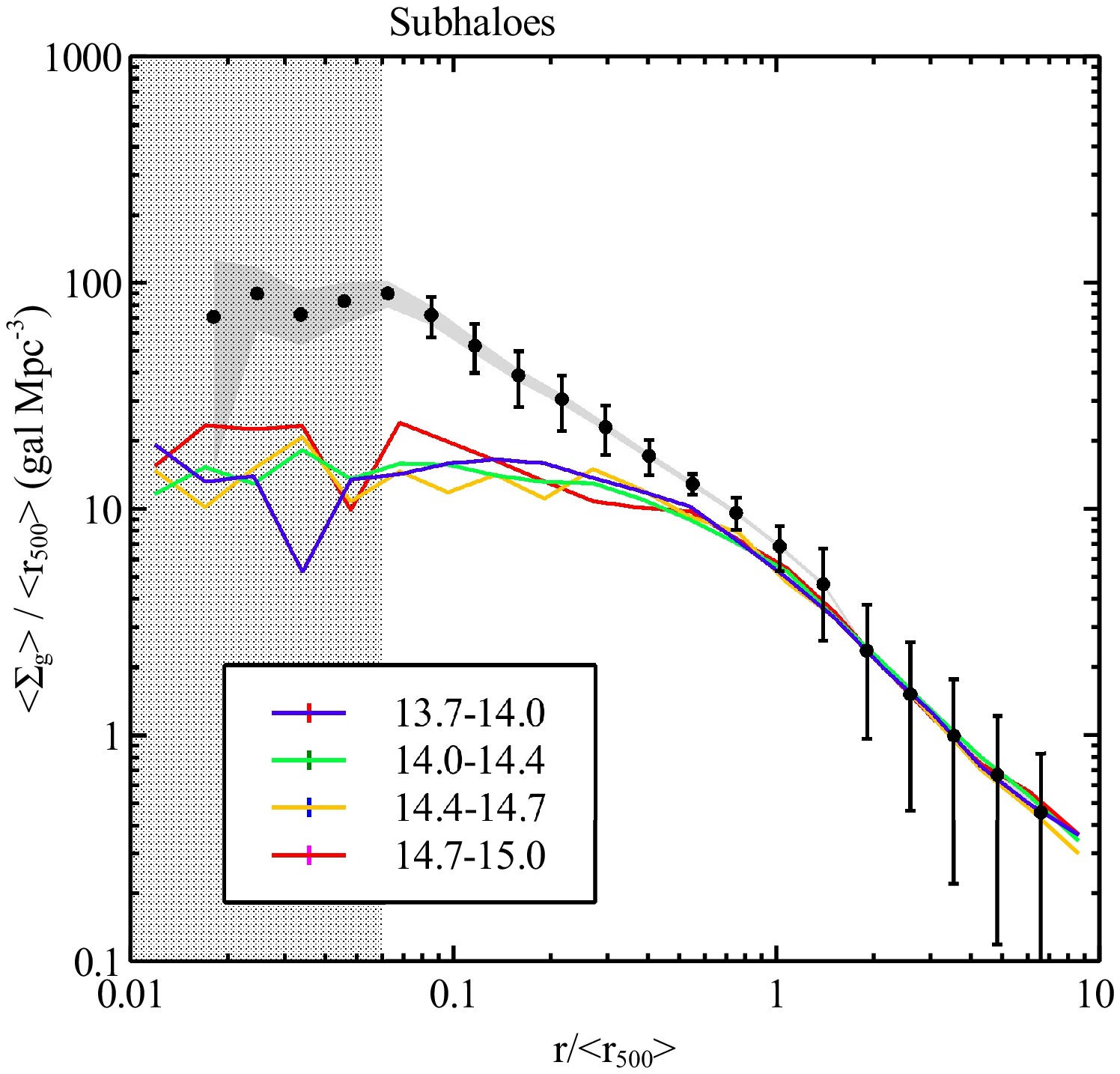}

\caption{Comparison between the observed satellite number density
  profile (solid black dots) and those from a series of semi-analytic
  models of galaxy formation (colored curves). The black solid line is
  a stack of all BCG clusters in the sample. The black error bars show the representative scatter between profiles within each stack, and the shaded grey region s hows the mean error bar due to Poisson scatter as a function of radius. The shaded (dotted) regions represent areas of
  incompleteness due to BCG obscuration. Panels
  (a),(b), and (c) correspond to the semi-analytic model satellites
  profiles of \citet{2006MNRAS.370..645B}, \citet{2008MNRAS.389.1619F}
  and \citet{2007MNRAS.375....2D} respectively. Panel (d) shows the
  model satellites profiles of \citet{2007MNRAS.375....2D} when only
  satellites with existing dark matter subhaloes are included.}
\label{cap:semianalytics}
\end{figure*}


\section{Semi-Analytic models}\label{sec:sams}

We can attempt to shed some light on the observed behaviour of the
satellites in clusters (Section \ref{sec:results}), by
comparing our results with predictions from semi-analytic models of
galaxy formation. In particular, we extract projected number density
profiles for the \citet{2006MNRAS.370..645B},
\citet{2008MNRAS.389.1619F} and \citet{2007MNRAS.375....2D} models
using the Millennium Simulation SQL
database\footnote{http://virgo.dur.ac.uk/}.  We select all
clusters from $z=0.32$ snapshot for which $\log M_{500} \ge
13.7$, yielding a sample of over 3000 simulated clusters.
To mimic the procedure used to derive the observed profiles, we
extract galaxies within a 5 Mpc aperture centered on the position of
the most bound particle (which should correspond to the location of
the BCG), perform a statistical background subtraction using a
randomly placed 5 Mpc aperture placed, and retain only galaxies with a
absolute rest-frame r-band magnitude $> -20.5$.

In Fig \ref{cap:semianalytics} we show the stacked satellite profiles
in four mass bins (the same ones as in Fig. \ref{cap:stackscaled}) for
the 3 models, along with a single stacked observational profile
comprised of all the clusters in the sample\footnote{The limited
  dependence of observed satellite concentration on halo mass in Fig
  \ref{cap:massconc}, implies that we can reasonably combine all
  clusters in the sample into a single universal
  observational profile for comparison with the models.}.  We will
discuss the bottom right panel below.

Underlying the \citet{2006MNRAS.370..645B},
\citet{2008MNRAS.389.1619F} and \citet{2007MNRAS.375....2D} models are
the same friends-of-friends (FoF) and substructure catalogs, which
were produced by running the \subfind\ algorithm (Springel et
al.\ 2001) on the Millennium Simulation.  There are some slight
differences in the merger trees constructed from these catalogs by the
Durham and Munich groups and significant differences in the
implementation of baryonic physics (such as radiative cooling, stellar
evolution, supernova feedback, AGN feedback and so on).  In principle
the models could therefore produce quite different results.  However,
it should be borne in mind that all three models have been tuned to
match the global galaxy luminosity function.  What this tuning
effectively achieves is to assign approximately the correct amount of
luminosity/stellar mass to the central galaxy of a given dark matter
halo (i.e., similar to what is achieved by direct `abundance
matching', but without explicitly imposing a monatonic relation
between luminosity/stellar mass and halo mass or a constant amount of
scatter at fixed halo mass).  Our expectation, therefore, is that all
three models ought to be similar at large radii and should, at least roughly, reproduce the observed {\it unscaled} satellite profiles there, where the
effects of tidal evolution and dynamical friction should be minimal.  Note that the model and observed {\it scaled} radial profiles will only match in normalisation at large radii if our estimates of $r_{500}$ are accurate, since we are using empirical estimates of $r_{500}$ for the observed clusters and the {\it true} values of $r_{500}$ for the
simulated systems.  As can be seen in Fig \ref{cap:semianalytics} all three models
have similar behaviour at large radii and approximately match the
observed profiles both in terms of shape and (encouragingly) normalisation.  There is also a lack of a strong halo mass
dependence in shape/normalisation of the profiles from the models, exactly as
observed.  

Within $r_{500}$ differences between the models become apparent.  In
particular, the \citet{2007MNRAS.375....2D} profiles become noticeably
flatter, in rough accordance with the observed profiles, whereas the
\citet{2006MNRAS.370..645B} and \citet{2008MNRAS.389.1619F} profiles
show a smaller degree of flattening.  At first sight, this may seem
suprising, as all three models use the FoF and substructure catalogs.
But note the difference between the models originates from differences
in the treatment of satellite galaxies when the subhalo to which that
galaxy belonged is no longer identified by \subfind\ .  Subhaloes can
be `lost' either because they have been completely tidally disrupted
or because the substructure finding algorithm has failed to find them.
In Fig \ref{cap:semianalytics} \emph{(d)} we show the profiles derived
from the \citet{2007MNRAS.375....2D} model when only satellite
galaxies with still {\it existing} subhaloes are included.  This is
equivalent to assuming that whenever a subhalo is lost the galaxy
within it is completely destroyed.  Here we see a very strong break in
the profiles within $r_{500}$, which is in obvious discord with the
observations.  The interesting implication of this experiment is that
tidal disruption of satellites galaxies is much less efficient than
the disruption of their surrounding dark matter subhalos\footnote{Here
  we are making the assumption that the substructure finders are able
  to find most self-gravitating dark matter haloes.}.

In all three models when a satellite galaxy loses its dark matter
subhalo the position of the satellite galaxy is assigned by using the
current position of the most bound dark matter particle of the subhalo
at the time the subhalo was last identified.  This procedure cannot be
followed indefinitely, of course, as no merging would take place with
the central galaxy, since a single dark matter particle does not
experience dynamical friction and therefore will not sink to the
center.  Therefore, what is done is to calculate a dynamical friction
merger timescale, $\tau_{\rm merge}$, for the satellite galaxy.  This
timescale is calculated differently in the Durham and Munich models.
The Durham models use a timescale that is proportional to the original
formation of \citet{1943ApJ....97..255C} (see \citealt{2000MNRAS.319..168C}
for details) and depends on the main halo mass, the mass of the
satellite, and orbital energy and angular momentum of the satellite.
We note here that the mass of the satellite that is used is the total
mass (gas+stars+dark matter) at {\it the time of virial crossing} of
the main halo, rather than the time the subhalo was last identified.
Also, the distribution of initial orbital parameters of infalling
satellites (at virial crossing) is adopted from a fit to the
cosmological simualtions of \citet{1997MNRAS.290..411T} rather than
using the orbital parameters of the satellites in the Millennium
Simulation (we note, however, that a newer version of the GALFORM code
allows for this possibility).  The constant of proportionality, which
is of order unity, is treated as a free parameter and varied to
improve the match to the break in the galaxy luminosity function.
Both the \citet{2006MNRAS.370..645B} and \citet{2008MNRAS.389.1619F}
models use the same constant of proportionality.  The Munich model, by
contrast, uses the properties of the satellite (i.e., its mass and
distance from the center) at the time its subhalo was last identified,
which should be better, but their timescale calculation lacks any
dependence on the orbital parameters, which should be relevant.  There
is also a constant of proportionality in the merger timescale in the
Munich model, which \citet{2007MNRAS.375....2D} have tuned to improve
the match to the break in the luminosity function.

Once a satellite galaxy has been in orbit for a time equal to the merger
timescale\footnote{Note the `clock' starts ticking at different times
  for the Durham and Munich models.  It starts at virial crossing for
  the Durham models and at the time the subhalo was last identified
  for the Munich model.} the satellite is removed from the catalog and
its mass is added to the central galaxy.  Note that none of the
models take into account mass loss due to tidal forces over the course
of this merger timescale.

The larger degree of flattening of the profiles from the
\citet{2007MNRAS.375....2D} model would suggest a generally shorter
merger timescale than that which is being adopted in the
\citet{2006MNRAS.370..645B} and \citet{2008MNRAS.389.1619F}.  Why this
is the case is difficult to ascertain, given the large differences in
the way in which these studies calculate this timescale.  Recently, \citet{2008ApJ...675.1095J}
and \citet{2008MNRAS.383...93B} used detailed
numerical simulations to show that the current implementations of the
dynamical friction merger timescale in the semi-analytic models do not
accurately reproduce the true merger timescale from the simulations,
in that they tend to underestimate the timescale for low mass
satellites and overestimate it for massive satellites.  This is likely
due to a failure of a number of assumptions, including the neglect of
mass loss (see \citet{2010PhR...495...33B} for further discussion).

In reality, the flattening of the satellites radial profiles is likely
to be due to a combination of merging with the central galaxy and
tidal mass loss of the satellites as they orbit about cluster.
The fact that galaxy clusters contain detectable amounts of
intra-cluster light (ICL) (see, e.g., \citealt{2005ApJ...618..195G};
\citealt{2005MNRAS.358..949Z}) demonstrates that the latter mechanism
has to be taking place as well.  Disentangling these two processes
relies upon understanding the complex interplay between the satellite
profiles, the BCG luminosities and the ICL. In a forthcoming study (Koposov et al, in
preparation) we will present self-consistent measure of ICL for the cluster sample presented here.  In principle this should allow us to take a large jump towards understanding the complicated interplay
between tidal stripping and satellite merging.

\section{Conclusions}

We have used the power of both spectroscopic and photometric datasets
published as part of the SDSS DR7 to measure the satellite profiles of clusters with Luminous Red Galaxies (LRGs) at
their centers. We take advantage of the large volume of the SDSS to
probe a substantial range in halo mass ($10^{13.7} < M_{500} \leq
10^{15.0} \ M_{\odot}$), and utilize the SDSS spectroscopic BCG sample
to probe clusters out to a reasonably high redshift ($z=0.4$).

Using the ``anchor'' sample of SDSS clusters with accurate X-ray
temperature measurements we link the optical richness with the
cluster's extent and total mass $r_{500}, m_{500}$. Armed with this
robust measurement of halo mass, we have constructed a large catalogue
of $\sim 20\,000$ BCG clusters, which is highly pure and
complete at halo masses of $M_{500}\gtrsim10^{14} \ M_{\odot}$. The
estimates of cluster mass are used to partition clusters in four bins
according to the total amount of dark matter. In each mass bin, high
signal-to-noise satellite profile stacks are produced. These reveal
that accross all masses, the satellites are systematically less
concentrated than the dark matter, roughly by a factor of two.  Interestingly, in spite of the difference in shape between the galaxy and DM radial distributions, both exhibit a high degree of self-similarity (i.e. no strong change in shape of the profile as a function of mass).
We find
a strong evolution in the concentration of the satellite profile as a
function of satellite brightness and colour and discuss physical mechanisms to explain this observed behaviour.  We find only very weak dependencies of the satellite radial distribution on cluster redshift (over the range 0.15-0.4) or BCG luminosity.

We made a self-consistent comparison with a number of recent semi-analytic
models of galaxy formation.  We showed that: (1) beyond approx. $0.3 r_{500}$ current models are able to reproduce both the shape and normalisation of the satellite profiles; and (2) within $0.3 r_{500}$ the predicted profiles are sensitive to the details of the satellite-BCG merger timescale calculation.  The former is a direct result of the models being tuned to match the global galaxy luminosity function combined with the assumption that the satellite galaxies do not suffer significant tidal stripping, even though their surrounding DM haloes can be removed through this process. The deviation within $\sim$0.3 r500 implies that the semi-analytic models have too long of a merger timescale and/or tidal stripping of the stellar component of galaxies, prior to merging with the BCG, is important (the effects of tidal stripping are not taken into account in the models
that we have considered).  The fact that groups and clusters demonstrably have non-neglibible stellar mass in a diffuse component (the ICL) indeed suggests that the latter process is relevant. In a forthcoming study we will present stacked measurements of the ICL for the cluster sample presented here.  When combined with our measurements of the satellite population and the BCG it should be possible to gain a more complete understanding of the evolution/fate of the satellite population and could serve as a means to inform theoretical models on the efficacy of the tidal stripping and merging processes.


\section*{Acknowledgements}

We thank the anonymous referee for helpful suggestions which improved the paper.
We would like to thank Ann Zabludoff, Dennis Zaritsky, Paul Hewett and Manda
Banerji for valuable discussions. JMB acknowledges the award of a STFC
research studentship, whilst VB acknowledges financial support from the
Royal Society. IGM acknowledges support from a Kavli Institute
Fellowship at the University of Cambridge and a STFC Advanced Fellowship at the University of Birmingham.  This work has made
extensive use of the \texttt{NumPy} and \texttt{SciPy} Numerical
Python packages and the \texttt{Veusz} plotting package. Funding for
the Sloan Digital Sky Survey (SDSS) has been provided by the Alfred
P. Sloan Foundation, the Participating Institutions, the National
Aeronautics and Space Administration, the National Science Foundation,
the US Department of Energy, the Japanese Monbukagakusho, and the Max
Planck Society. The SDSS website is http://www.sdss.org/.  The SDSS is
managed by the Astrophysical Research Consortium (ARC) for the
Participating Institutions. The Participating Institutions are The
University of Chicago, Fermilab, the Institute for Advanced Study, the
Japan Participation Group, The Johns Hopkins University, Los Alamos
National Laboratory, the Max Planck Institute for Astronomy (MPIA),
the Max Planck Institute for Astrophysics (MPA), New Mexico State
University, the University of Pittsburgh, Princeton University, the
United States Naval Observatory, and the University of Washington.

\label{lastpage}

\bibliographystyle{mn2e.bst}
\bibliography{biblio.bib}


\appendix
\section{Linear model fitting} \label{app:fit}
This describes the formalism from \citet{2010arXiv1008.4686H} for the fitting of a straight line model with slope $m$ and intercept $b$, to data with two dimensional uncertainties.

\noindent The orthogonal displacement $\Delta_i$ of each data point from the line is given by:

\begin{equation}
\Delta_i = \mathbf{v}^\mathrm{T} x_i y_i - b\,\cos\theta\;,
\end{equation}

\noindent where $(x_i,y_i)$ is the individual data point, and $\mathbf{v}$ is the unit vector orthogonal to the line given by:

\begin{equation}
\mathbf{v}
 = \frac{1}{\sqrt{1+m^2}}\,\left[\begin{array}{c}-m\\1\end{array}\right]
 = \left[\begin{array}{c}-\sin\theta\\\cos\theta\end{array}\right]\;.
\end{equation}

\noindent Each data points covariance matrix is given by:

\begin{equation}\label{eq:Sigma}
\Sigma_i^2 = \mathbf{v}^\mathrm{T}\,\mathbf{S}_i\,\mathbf{v}\;,
\end{equation}

\noindent where the covariance matrix of horizontal and vertical measurement uncertainties corresponds to:

\begin{equation}
\mathbf{S}_i \equiv \left[\begin{array}{cc}
\sigma_{xi}^2 & \sigma_{xyi} \\ \sigma_{xyi} & \sigma_{yi}^{2}
\end{array}\right] \quad  \;.
\end{equation}



\section{MaxBCG radial profiles}\label{app:maxbcg}
\begin{figure}
\includegraphics[width=\columnwidth]{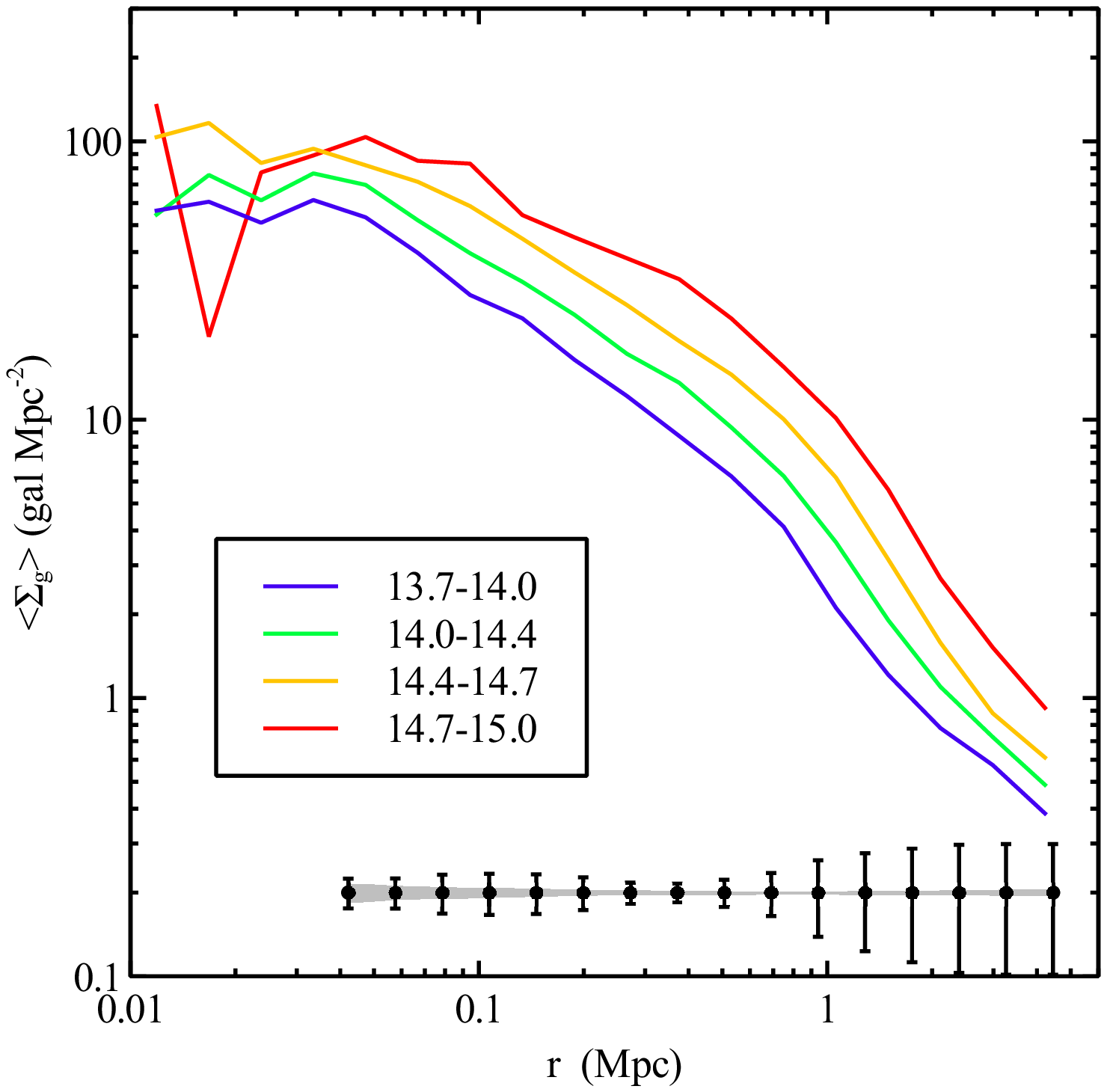}
\caption{The mean satellite number density profiles as a function of
  radius for stacks of MaxBCG cluster satellites in bins of halo
  mass. The black points show the representative mean error bar on the profiles as a function of radius. It is clear that the higher mass clusters have a larger
  number of satellites than the lower mass bins.}
\label{cap:stackraw_bcg}
\end{figure}

Fig. \ref{cap:stackraw_bcg} shows the raw unscaled satellite number
density profiles for the MaxBCG clusters \citep{2007ApJ...660..239K}, in
four mass bins. The profiles are very similar to those generated using
the cluster sample described in this paper (Fig. \ref{cap:stackraw}).

\section{Error modelling}\label{app:errmodel}

In order to estimate the scatter in number density profiles we perform the following analysis. As described in the text we are interested in
 the scatter of the scaled surface density: $\Sigma_{gal}={1}/{{\rm R}_{500}}\,\Sigma_{0,gal}$ as a function of log(R/R$_{500}$). Our data consist of
the measurements of the number of galaxies in circular annuli for different clusters and the corresponding measurements of the background galaxy density.

First we consider the expected number of galaxies in the $i$-th bin of log(R/R$_{500}$) for the galaxy cluster $j$:

\begin{equation}
N_{expected,i,j} = A_{i,j}\,(R_{500,j}\,\Sigma_{i,j}+\Sigma_{bg,j})
\end{equation}

\noindent where $A_{i,j}$ is the area of the $i$-th annulus for $j$-th cluster, $R_{500,j}$ is the $R_{500}$ of the $j$-th cluster,
$\Sigma_{i,j}$ is a scaled density of galaxies in the cluster
and $\Sigma_{bg,j}$ is the background galaxy density. In the simplest scenario we would assume that $\Sigma_{i,j} = \Sigma_{i} $ is constant
for all clusters, but in this case we will assume that the $\Sigma_{i,j}$ is Gaussian distributed with the mean $\Sigma_{i}$ and dispersion $S_i$.

Now we can write the likelihood of the parameters $(\Sigma_i,S_i,\Sigma_{bg,i},\Sigma_{i,j})$ given observed $N_{i,j}$ galaxies in the $i$-th bin of the $j$-th cluster and $N_{bg,i}$ background
galaxies observed in the random 5\,Mpc radius field:
\begin{eqnarray}
{\mathcal L_j}(\Sigma_i,S_i,\Sigma_{bg,i},\Sigma_{i,j}|N_{i,j},N_{bg,i}) &=& \nonumber \\  
P(N_{i,j}|N_{expected,i,j})\,P(\Sigma_{i,j}|\Sigma_{i},S_i)\,P(N_{bg,i}|\Sigma_{bg,i}) 
\end{eqnarray}

\noindent where the first term is Poisson probability of observing the $N_{i,j}$ objects given a certain density of galaxies (including background), the second term is the
probability of having $\Sigma_{i,j}$ given the mean density $\Sigma_{i}$ and scatter $S_{i}$, and the third term is essentially is Gaussian probability of the
background measurement from the random 5\,Mpc radius field. In our case we are not interested in the values of $(\Sigma_{i,j},\Sigma_{bg,i})$, so their are nuisance
parameters for us and we can marginalize over them. Then the likelihood becomes only a function of  $\Sigma_{i},S_i$:

\begin{equation}
{\mathcal L_{i,j}}(\Sigma_i,S_i,) = \int {\mathcal L_{i,j}}(\Sigma_i,S_i,\Sigma_{bg,i},\Sigma_{i,j})\, d \Sigma_{bg,i}\, d \Sigma_{i,j}\,.
\end{equation}

\noindent The full likelihood for the $i$-th bin of the profile is then obtained by multiplying the likelihoods for different clusters

\begin{equation}
{\mathcal L_{i}}(\Sigma_i,S_i)= \prod\limits_j{\mathcal L_{i,j}}(\Sigma_i,S_i)\,.
\end{equation}

\noindent We then maximize that likelihood with respect to the $(\Sigma_i,S_i)$ parameters for each radial bin, which gives as estimates of the mean and a scatter of the
density profiles. For the error estimates of the parameters we use Hessian matrices numerically evaluated at the best-fit parameters. 


\end{document}